\documentclass[preprint,12pt]{elsarticle}

\usepackage{amssymb}

\journal{Annals of Physics}
\usepackage{graphicx}
\usepackage{bm}

\usepackage{color}

\definecolor{refcol}{rgb}{0,0,.5}
\usepackage{microtype}
\usepackage[colorlinks,linkcolor=refcol,citecolor=refcol,urlcolor=refcol]{hyperref}

\definecolor{red}{rgb}{1,0,0}
\definecolor{blue}{rgb}{0,0,1}

\usepackage{amsmath}

\biboptions{comma,sort&compress}

\makeatletter
\newcommand{\cz}{
  \mathord{\mathpalette\vaggelis@z{z}}%
}

\newcommand{\cZ}{
 Z%
}

\newcommand{\vaggelis@z}[2]{%
  \sbox\z@{$\m@th#1#2$}%
  \ooalign{%
    $\m@th#1#2$\cr
    \hidewidth
    \vrule height \dimexpr.5\ht\z@+0.03ex\relax
           depth -\dimexpr.5\ht\z@-0.03ex\relax
           width .5\wd\z@
    \hidewidth\cr
  }%
  \vphantom{\box\z@}
}
\makeatother

\newcommand*{\scv}{z}

\begin{document}

\begin{frontmatter}

\title{Universal location of Yang-Lee edge singularity for a  one-component field theory in $1\le d \le 4$ }

\author[a,b]{Fabian Rennecke}
    \address[a]{Institute for Theoretical Physics, Justus Liebig University Giessen, Heinrich-Buff-Ring 16, 35392 Giessen, Germany}
   \address[b]{Helmholtz Research Academy Hesse for FAIR (HFHF), Campus Giessen, 35392 Giessen, Germany}
 \ead{fabian.rennecke@theo.physik.uni-giessen.de}

\author[c,d]{Vladimir V. Skokov}
    \address[c]{Department of Physics, North Carolina State University, Raleigh, NC 27695, USA}
    \address[d]{RIKEN BNL Research Center, Brookhaven National Laboratory, Upton, NY 11973, USA}
    \ead{VSkokov@ncsu.edu} 

\begin{abstract}
We determine    
the universal location of the Yang-Lee edge singularity 
in the entire relevant domain of spatial dimensions $1\le d \le 4$
for the Ising universality class. To that end, we present analytical results for $d=1,2,4$ and near four dimensions. 
For $d=3$ and a set of fractional dimensions, we perform numerical calculations using a systematic Functional Renormalization Group approach. 
\end{abstract}

\date{\today}
\end{frontmatter}


\newpage 

\tableofcontents

\section{Introduction}
\label{sec:intro}

Lee and Yang   demonstrated an intimate connection between the analytical structure of the equation of state in the complex plane of the thermodynamic parameters and  the phase structure of the system \cite{PhysRev.87.404,PhysRev.87.410}, see also Refs.~\cite{PhysRevLett.27.1439,Fisher:1978pf}.
To introduce the key concepts of interest it is best to continue  in a well-known Ising model. 
Above the Curie temperature, the model demonstrates a smooth transition (commonly referred to as  ``crossover'' in the finite temperature QCD community) as a function of the external magnetic field, $h$. The Lee-Yang theorem infers that going to the full complex plane of $h$ reveals  the presence of  two branch points (and associated branch cuts)  at purely imaginary values  $h = \pm i h_c $ (for a comprehensive review see Refs.~\cite{bena2005statistical}). When the Curie temperature is approached from above, the branch points pinch the real $h$-axis, leading to the emergence of the  physical critical point~\cite{Itzykson:1983gb}. These branch points are known as the Yang-Lee Edge (YLE) singularities. The described phase structure  exposes the fact that the conventional critical point  has a smaller codimension then the YLE singularities. In other words, to tune in to the critical point, typically two parameters have to be adjusted: $T$ and $h$; in contrast, for the YLE, it is sufficient to adjust only one parameter $h$ (albeit imaginary). For this reason, Michael Fisher~\cite{Fisher:1978pf,Fisher:1982yc} refereed to   the YLE singularity as to a {\it proto-critical} point (cf.\ critical and multi-critical points). Since the YLE singularity has only one relevant variable, it also has  only one truly independent critical exponent -- the edge critical exponent $\sigma$ (see Ref.~\cite{PhysRevLett.27.1439} for a first study of $\sigma$). It characterizes the behaviour of the magnetization ($M$) in the vicinity of the YLE:  $M\sim M_{\rm c} + (h - h_c)^{\sigma}$. In three dimensions, the value of the edge critical exponent is not known analytically; its numerical value has been refined in many studies, see e.g. Ref.~\cite{Gliozzi:2014jsa,Gracey:2015tta,Borinsky:2021jdb,An:2016lni, Zambelli:2016cbw}.  

It is well known, that the critical properties 
of the system  (e.g.\ Ising model near the critical point) can be  quantified by a universal function of one variable (see e.g.\ Refs~\cite{Cardy:1996xt,Amit:1984ms}), e.g.\ the magnetic equation of state $M=h^{1/\delta} f_G(\scv)$, where $\scv$ is the scaling variable $\scv=t h^{- 1/\Delta}$ with  $\Delta=\beta\delta$.  $\beta$ and $\delta$ are universal critical exponents and $f_G(\scv)$ is the universal scaling function.

The presence of the branch points above the Curie temperature implies that the scaling function has the YLE edge singularities in the
complex plane of the scaling variable $\cz$. The corresponding value $\scv_c$ (and its complex conjugate) is a {\it universal} number. Our goal is to determine it for the Ising universality class. 

Why is the universal location 
of the YLE  singularity important? There are multiple reasons. 
First, in general, any universal number is of importance to characterize the universality class. 
Second, the YLE  is a singularity of the scaling equation of state and it is located at finite values of the scaling variable $\scv$. Thus the singularity  determines many properties of the scaling function for real values of the parameters. It limits the radius of convergence and through the Darboux theorem (see e.g.~Ref.~\cite{henrici1991applied}), the asymptotically high order coefficients of the Taylor expansion of $f_G(\scv)$ about any finite argument are largely determined by $\scv_c$ and $\sigma$.  Finally, there is a practical reason. It  stems from application to specific systems  in the same universality class. For example,  a universal input complemented by the non-universal mapping of parameters can facilitate reconstruction of the QCD phase diagram at finite baryon chemical potential. 
Due to the sign problem, direct lattice QCD calculations can not be effectively performed at non-zero baryon chemical potential/density $\mu\ne 0$. We thus are limited to working with indirect methods, for example Taylor series expansion of the pressure about $\mu=0$. 
This series expansion has a finite radius of convergence which, in many feasible scenarios,  is defined by the YLE edge singularity $\scv_c$ (as was first recognized in Ref.~\cite{Stephanov:2006dn}) and a non-universal map of $T$, $\mu$, and $m$ to the scaling variable $\scv$. We refer an interested reader to a recent studies of Refs.~\cite{Mukherjee:2019eou,Basar:2021gyi,Basar:2021hdf}.   

Why was  not  the location determined long ago? The answer is two-fold.  On one hand, as we also will show in this paper, the commonly used framework of the $\varepsilon$-expansion near four dimensions  leads to non-perturbative contributions beyond the leading order, which significantly limits it predictive power for $d=3$. 
This is related to the fact that the Wilson-Fisher fixed point at vanishing $h$ has an upper critical dimension of four, while the YLE fixed point at imaginary $h$ has upper critical dimension six~\cite{Fisher:1978pf}.
On the other hand, the lattice methods based on importance sampling prohibit direct computations at imaginary values of the magnetic field due to  the sign problem~\footnote{In a finite volume, 
the Yang-Lee edge singularity and the associated branch cuts are replaced by the Lee-Yang zeroes. Extracting the zeroes is also a challenging problem using lattice simulations, see e.g. Refs.~\cite{Giordano:2019gev,Attanasio:2021tio,Nicotra:2021ijp,Dimopoulos:2021vrk,Ejiri:2005ts}. A quantum computer might be a way to tackle this problem, see e.g.\ Refs.~\cite{Wei:2016oyn,francis2021manybody} and references therein.}; thus rendering the problem of extracting the location of the YLE singularity practically impossible.

The FRG approach allows to circumvent both issues, as it does not rely on perturbative expansion as well as it does not suffer from a sign problem.
As far as the Ising universality class is concerned,   in this manuscript, we significantly expand the result reported before, see  Ref.~\cite{Connelly:2020gwa}. First, we provide (semi) analytical results for $d=1,2$ and in the vicinity of four dimensions $d=4-\varepsilon$. Second, we perform numerical calculations using the FRG 
{at next-to-leading order in the derivative expansion \cite{Morris:1994ie, Litim:2001dt, Balog:2019rrg}.}
Finally, we put forward a novel approach which allows us to follow the FRG flow  directly to the YLE singularity in the infrared (IR) without using extrapolation procedure applied in Ref.~\cite{Connelly:2020gwa}. We believe this significantly improves the precision of the results by suppressing the systematic uncertainty.        

As we already noted, the FRG is a non-perturbative approach that circumvents the sign problem 
and that enables us to perform calculations at complex values of thermodynamic parameters. 
However, locating the YLE singularity in the FRG approach is numerically a much more challenging undertaking than computing the  critical exponents at a Wilson-Fisher or at YLE points. We briefly explain why. The practical implementation of the FRG boils down  to a set of integro-differential equations. To determine the critical exponents it is sufficient to find a fixed point of the set; thus the problem reduces to a set of algebraic equation involving integrals. It can be tackled with the Newton-Raphson method and a fast integrator, e.g.\ a fixed order Gauss–Kronrod quadrature. In contrast to this, locating YLE requires solving the actual set of the integro-differential equations. In order to make this problem manageable on the current hardware, we use a specific form of the  cut-off function, see Sec.~\ref{Sec:FRG}, that allows for analytical evaluation of the integrals and simplifying the problem to the set of differential equations. The disadvantage of this choice of the regulator is that it limits the order of the derivative expansion we can work at. This in combination with the computational constraint  does not let us to match the precision of our calculation to that of the critical exponents in Refs.~\cite{Balog:2019rrg, DePolsi:2020pjk}.

This manuscript is organized as follows. In Sec.~\ref{Sec:Scaling}, we review the main concepts of interest, such as the scaling equations, critical amplitudes and exponents. In Sec.~\ref{Sec:Analytic}, we then turn to extracting the location of the YLE singularity in $d=1, 2$ and at/near four dimensions using (semi)analytic methods. 
At three dimensions, we have to resort to numerical methods. The method of our choice is the FRG approach. We introduce it and present the results in Sec.~\ref{Sec:FRG}. We finally end with Conclusions.   

\section{Scaling equation, critical amplitudes and  exponents}
\label{Sec:Scaling}
In this section, we  will provide a brief review of the critical statics. We limit the scope of our discussion only to the concepts of relevance for what follows. For a comprehensive review, we refer the reader to Refs.~\cite{Amit:1984ms,Zinn-Justin:2002ecy,Vasilev:2004yr}.  

In the vicinity of a critical point, the thermodynamic quantities  exhibit power-law behaviour. It is characterized by a set of universal critical exponents and, in general, non-universal amplitudes (for a review of universal amplitude ratios, see e.g. Ref.~\cite{Zinn-Justin:2002ecy}). In this paper, we study single component field theory with $\cZ(2)$ symmetry. Thus we have to consider only two relevant perturbations from the critical point: the
temperature-like variable which preserves the symmetries of the system, $t = (T -T_c)/T_c$, and the explicit symmetry breaking external field, $H$. 
At zero external magnetic field, the order parameter (e.g. the magnetization in the Ising model) in the spontaneously broken phase ($t<0$) is proportional to the reduced temperature to power $\beta$:
    \begin{align}
        M = B (-t)^\beta \,.
    \end{align}
    Here $B$ is a non-universal amplitude. 
    
    In the symmetric phase, $t>0$,   the order parameter susceptibility is divergent with the exponent $-\gamma$
    \begin{align}
        \label{Eq:Chi}
        \chi(t,H=0) = \frac{ \partial M} {\partial H}  = C_+ t^{-\gamma}\,.
    \end{align}
    A similar relation, albeit with a different amplitude $C_-$, is true in the spontaneously broken phase. 
    
  On the critical isotherm $t=0$, the order parameter is a power-law in the magnetic field  
    \begin{align}
    \label{Eq:M_tc}
        M = B_c H^{1/\delta}  \,.
    \end{align}
These three relations are of the most significance for the discussion in this section. For completeness we also consider the behaviour of the order parameter two-point correlation function $G(|x|)$ 
near and at the critical point. Near $T_c$ and at zero $H$,
\begin{align}
    G(|x|) = |x|^{-(d-2)} g(|x|/\xi), 
\end{align}
where $\xi$ is the only relevant scale, called the correlation length. It diverges at the critical point as  
\begin{align}
    \xi \propto |t|^{-\nu}\,.   
\end{align}
Finally, exactly at the critical   temperature and at zero magnetic field, the correlation function has a power law dependence on the  distance:
\begin{align}
    G(|x|) \sim |x| ^ {-(d-2+\eta)}\,.  
\end{align}
The critical exponent $\eta$ (also known as the anomalous dimension)  will appear later when we discuss the  Functional Renormalization Group approach.  
The introduced exponents are not all independent (only two critical exponents, $y_t$ and $y_h$ introduced below are) as there are multiple scaling relations tying them together. We will make use of the following: 
\begin{align}\label{Eq:nusc}
    \nu d &= \beta (\delta+1) = 2 \beta + \gamma\,, \\
    \label{Eq:etasc}
    2 - \eta &= d \frac{ \delta -1}{\delta+1}\,.
\end{align}

We now come back to the order parameter dependence on $t$ and $H$.   Usually one performs  rescaling of the relevant parameters $\bar t = \mu_t t$ and $h = \mu_h H$ by introducing the so-called metric factors $\mu_t$ and $\mu_h$  
\begin{align}
        \mu_t^\beta &=  B, \\
        \mu_h^{1/\delta} &= B_c\,  
\end{align}
in such a way as to absorb the amplitudes in the relations of the order parameter to the relevant perturbations:
\begin{align}
        \label{Eq:Mt}
        M(\bar t,h=0) &=  (-\bar t)^\beta, \\
        \label{Eq:Mh}
        M(\bar t =0,h) &= h^{1/\delta}\,.  
\end{align}
After this redefinition of the variables  the susceptibility at zero magnetic field and $t>0$ is given by  
\begin{align}
\label{Eq:susc}
    \frac{\partial M}{\partial h} = R_\chi (\bar t)^{-\gamma}, 
\end{align}
where $R_\chi$ is a universal amplitude ratio
\begin{align}
    R_\chi = \frac{C^+ B^{\delta -1 }}{B_c^\delta}\,.
\end{align}

\begin{figure}
    \centering
    \includegraphics[width=0.48\textwidth]{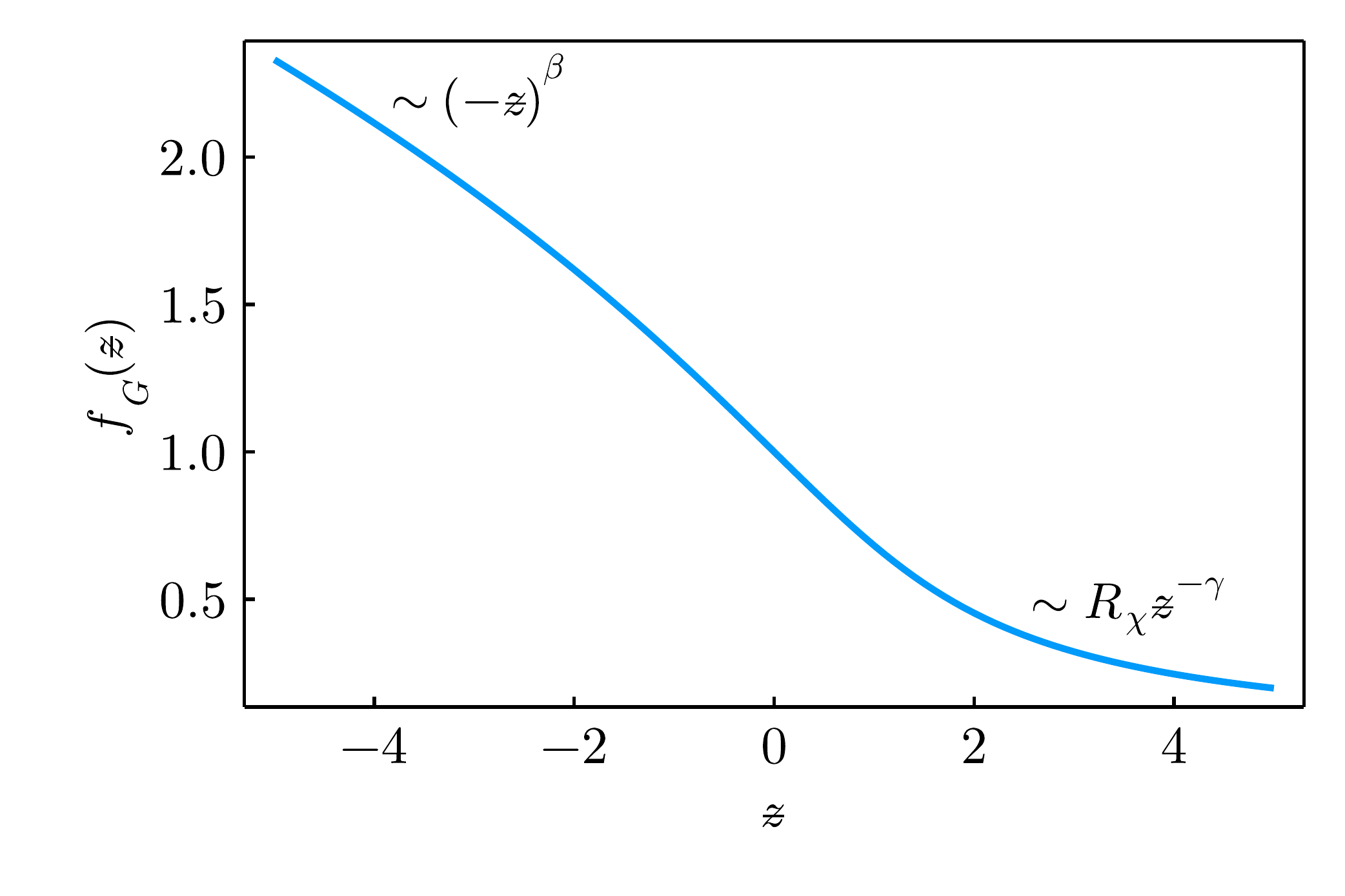}
    \includegraphics[width=0.48\textwidth]{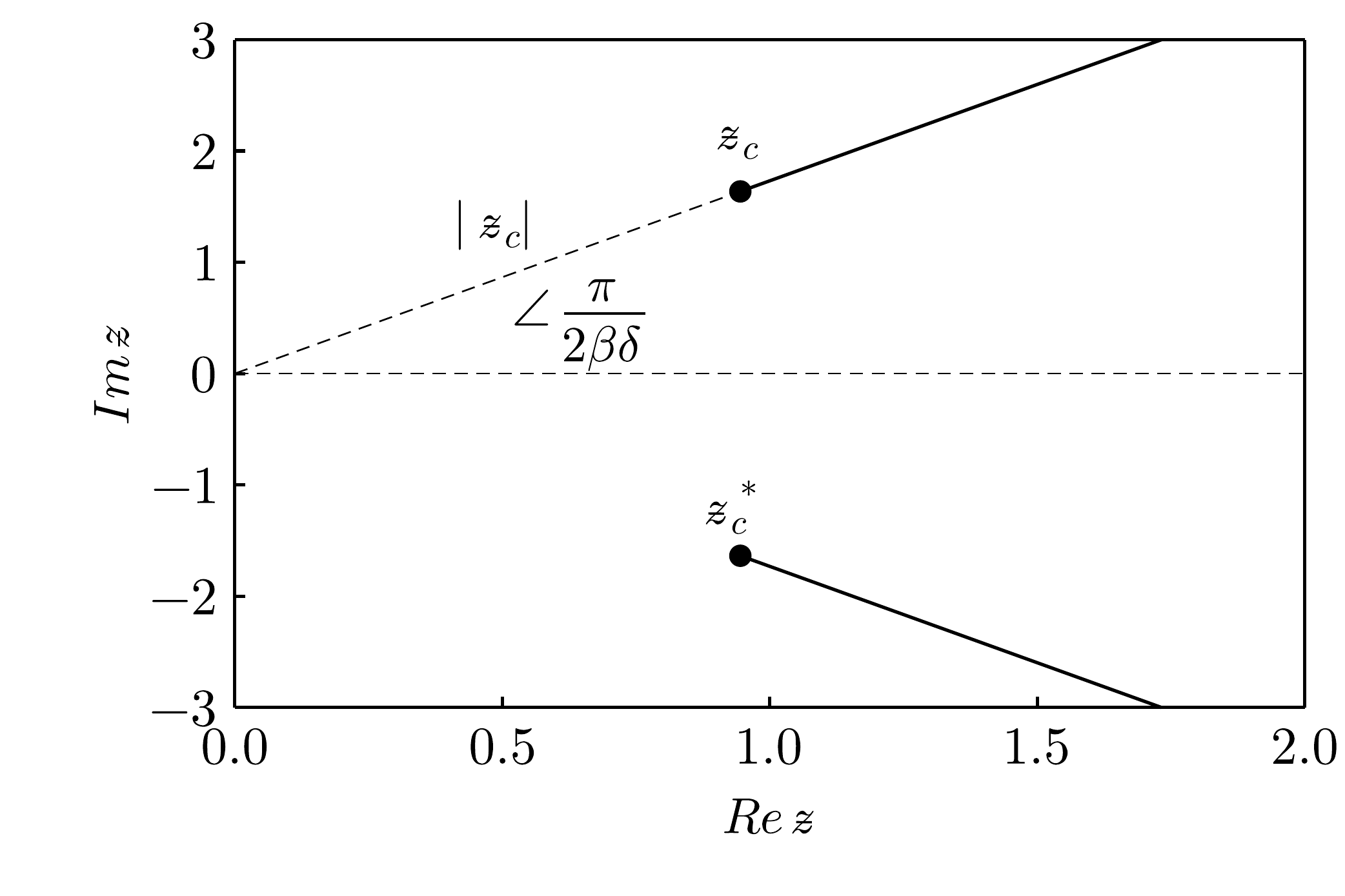}
    \caption{Left panel: An illustration for the magnetic equation of state $f_G(\scv)$ in mean-field approximation for real values of the argument $\scv$ (for the Ising universality class in two and three  dimensions, see Ref.~\cite{Caselle:2000nn,Fonseca:2001dc} and  Ref.~\cite{PhysRevE.65.066127} respectively). 
    Right panel: the analytical structure of the function  $f_G(\scv)$ in the complex $\scv$ plane. The YLE singularity (its complex conjugate) are located at $\scv_c$ ($\scv_c^*$). The lines extending from the singularities depict the branch cuts.}
    \label{fig:fg}
\end{figure}

The renormalization group equations allow us to generalize the above scaling laws to account for deviations in both relevant direction at the same time. Consider the free energy 
\begin{equation}
    f_s(t,h) = b^{-d} f( \bar t\, b^{y_t},  h  b^{y_h}),  
\end{equation}
where $b$ is a positive, arbitrary RG scale. 
The exponents $y_t$ and $y_h$ are the RG eigenvalues.
They are trivially related to the critical exponents defined above,  $y_t = 1/\nu$ and $y_h=\beta \delta/ \nu$, as one can easily establish. 
Taking a derivative with respect to $h$, we get magnetic equation of state
\begin{equation}
   M(t,h) = b^{-d+y_h} f^{(0,1)}( \bar t\, b^{y_t},  h  b^{y_h}),  
\end{equation}
Choosing $b$ appropriately one can eliminate one of the variables.
Usually the convenient choice is 
to set $ h  b^{y_h} =1$:
\begin{equation}
   M(t,h) = h^{(d-y_h)/y_h} f^{(0,1)}( \bar t\, /h^{y_t/y_h},  1) \equiv h^{(d-y_h)/y_h} f_G( \bar t\, /h^{y_t/y_h})\,.
\end{equation}
Substituting $y_h$ and $y_t$ we obtain 
\begin{equation}\label{eq:fG}
   M(t,h) = h^{1/\delta} f_G( \scv = \bar t\, / h^{1/\Delta})
\end{equation}
where we took into account the relation between critical exponents \eqref{Eq:nusc}. We also introduced the so-called gap critical exponent, $\Delta=\beta\delta$. 
The function $f_G$ is a function of one variable; it  encodes most of the critical statics. It has to satisfy the normalization conditions 
\begin{align}
\label{Eq:Normalization}
        f_G(0)  &=  1, \\
        \lim_{\scv\to - \infty} f_G(\scv) &\to (-\scv)^\beta\,. 
\end{align}
The former is rather straightforward to prove  as on the critical isotherm, one has  to recover Eq.~\eqref{Eq:Mh}. The latter is the consequence of   Eq.~\eqref{Eq:Mt}. Indeed, at vanishing $h$ and $t<0$ we have 
\begin{align}
    M(t<0, h\to 0) = \lim_{h\to 0} h^{1/\delta} (-\scv)^\beta = 
    (-\bar t)^\beta \,.
\end{align}
Another useful property of the scaling function $f_G$  
\begin{align}
\label{Eq:fGLargez}
        \lim_{\scv\to \infty} f_G(\scv) \to R_\chi \cdot(\scv)^{-\gamma} \, 
\end{align}
follows from Eq.~\eqref{Eq:susc}. We illustrate the magnetic equation of state by its mean-field approximation in right panel of Fig.~\ref{fig:fg}. The unassuming behaviour of this function for real values of the the argument does not give away the presence of the YLE singularities in the complex plane, see left panel of Fig.~\ref{fig:fg}. 

For the specific FRG calculation which follows in Sec.~\ref{Sec:FRG}, studies  in the spontaneously broken phase are rather challenging. This prevents us from computing the amplitude $B$ and the critical exponent $\beta$. While the latter can be easily obtained from the scaling relations, the former can be expressed in terms of the universal ratio $R_\chi$ and amplitudes $C_+$ and  $B_c$. 
It is thus  convenient to  introduce the ratio 
\begin{align}
    \label{Eq:zeta}
    \zeta = \frac{\scv}{R_\chi^{1/\gamma}}\,. 
\end{align}
Straightforward algebra shows that this ratio 
\begin{align}
\label{Eq:zetaAmpl}
    \zeta = \left( \frac{ B_c} { C_+ }  \right)^{1/\gamma} \frac{ t  } { H^{1/\Delta} }  
\end{align}
does not require the knowledge of the amplitude $B$. From the perspective of the scaling equation of state, this is  merely a redefinition of the argument of $f_G$.

\section{(Semi-)analytical results for  location of Yang Lee Edge singularity} 
\label{Sec:Analytic}
The location of the YLE singularity can be determined by finding zeros of the inverse  magnetic field susceptibility, that is
\begin{align}
    \label{Eq:YLElocgen}
    \frac{1} {f'_G(\scv_c)} =  \frac{d \cz(f^c_G)}{ d f^c_G} = 0  
\end{align}
where in the last equality we treated $\cz$ as a function of $f_G$ and defined $f^c_G = f_G(\scv_c)$. 
Note that the function $f_G(\scv)$ is universal and therefore the location of the  YLE singularity, $\scv_c$, determined by Eq.~\eqref{Eq:YLElocgen}, is universal. 
Moreover the corresponding $\zeta_c$ is a universal number as it is defined by the universal quantities only ($\scv_c, R_\chi$ and $\gamma$), see Eq.~\eqref{Eq:zeta}. 

From the Lee Yang theorem, implying that the YLE singularity is located at purely imaginary $H$, we can immediately conclude that the  arguments of both complex numbers $\scv_c$ and $\zeta_c$ is
\begin{equation}
\label{Eq:Arg}
    {\rm Arg} \, \scv_c =  {\rm Arg}\,  \zeta_c  = -\frac{1}{\Delta}
     {\rm Arg}\,  H_c =  \pm  \frac{\pi}{2 \Delta}\,.
\end{equation}
That is, the argument of the universal location is fully determined by the Ising critical exponents $\Delta = \beta \delta$ and does not require any extraneous knowledge (see Fig.~\ref{fig:fg} for an illustration). It is quite different for the absolute value $\zeta_c$  (or $|\scv_c|$), as it represents a completely independent universal number non-derivable from  a set of known critical exponents and amplitude ratios. Determining the  $\zeta_c$ is the goal of this work.

In general, since the analytic expression for $f_G$ is unknown, one rarely can find  the YLE singularity in this direct way.  Nevertheless there are a few exact results available for $f_G$ in certain limits. In the remaining part of this section,  we will consider (semi-)analytic results for  the Ising universality class, for $d=1$, $d=2$, $d=4-\epsilon$ and finally  $d=4$. We start with the latter.  

\subsection{Mean-field}
\label{Sec:MF}
Above  the upper critical dimension (4 for the Ising universality class)  the critical fluctuations can be neglected. In this case the mean-field approximation provides an accurate description of the critical behaviour.
For one component order parameter, the free energy is 
\begin{align}
    f = \frac{1}{2} t \phi^2 + \frac{\lambda}{4} \phi^4 - H \phi   \,.
\end{align}
The minimum  of the free energy defines the expectation value of the order parameter, or the magnetization, $M = \langle \phi \rangle$: 
\begin{align}
\label{Eq:MF}
    t M + \lambda M^3 - H = 0.
\end{align}
Therefore at zero magnetic field and $t<0$ 
\begin{align}
 M = \left( - \frac{ t }{\lambda} \right)^{1/2}
\end{align}
while at zero $t$, 
\begin{align}
 M = \left(\frac{ H }{\lambda} \right)^{1/3}\,. 
\end{align}
These relation define the mean-field critical exponents $\beta =1/2$ and $\delta =3$ as well as the metric factors. The equation governing the corresponding scaling function $f_G$, defined in Eq.~\eqref{eq:fG}, follows from Eq.~\eqref{Eq:MF}:  
\begin{align}
    \label{Eq:fGMF}
    f_G (\scv + f^2_G) = 1 \,. 
\end{align}
In principle one can solve this equation analytically and determine $f_G(\cz)$. However for our purposes of locating the YLE singularity we can avoid this. 
Differentiating with respect to $f_G$, remembering that  $  \frac{d \scv(f_G)}{ d f_G} = 0 $ at the YLE, we obtain
\begin{align}
    \scv_c + 3 f^2_G(\scv_c) = 0  \,. 
\end{align}
The simplest way to proceed is by multiplying everything by $f_G$ and using the magnetic equation of state, 
\begin{align}
    f_G(\scv_c)\left ( \scv_c + 3 f^2_G(\scv_c) \right)  =  
    1 + 2 f^3_G(\scv_c) = 0, 
\end{align}
with the result 
\begin{align}
    f_G(\scv_c) = \frac{ e^{\frac{i (\pi + 2\pi n) }{3}} } { 2^{1/3} }\,,\qquad
    \scv_c = - 3  \frac{ e^{\frac{i  2 (\pi + 2\pi n) }{3}} } { 2^{2/3} }\,.
\end{align}
We thus found three candidates; only  the  complex conjugate pair corresponds to the YLE singularities, they are    
\begin{align}\label{Eq:z_c}
    &\scv_c =  e^{  \pm i \frac{ \pi } { 3} } \frac{ 3 } { 2^{2/3} }\,.
\end{align}
Therefore the absolute value is 
\begin{align}
    |\scv_c| =  \frac{ 3 } { 2^{2/3} }\approx 1.88988\,.
\end{align}
From the asymptotic behaviour of $f_G$ in Eq.~\eqref{Eq:fGMF}  at large positive $\cz$~\eqref{Eq:fGLargez}, one can read off $R_\chi=1$. Therefore, we simply have $\zeta_c = \scv_c$. 
Note that the argument of  $\scv_c$, see Eq.~\eqref{Eq:z_c}, is equal to $\pi/3$, as expected, c.f. Eq.~\eqref{Eq:Arg} with $\Delta = 3/2$. 

We note  that this prediction of the mean-field approximation is accurate for $d\ge 4$.  

\subsection{$\varepsilon$-expansion} 
\label{Sec:Epsilon}
The result obtained in the previous section can be generalized to $4-\varepsilon$ dimensions. 
Although for many universal critical quantities, the $\varepsilon$-expansion in combination with asymptotic series analysis techniques provides a well-defined systematic way to extract numerical values, the location of the YLE singularity is somewhat exceptional as was pointed out in Ref.~\cite{An:2017brc}. As we demonstrate below, $\scv_c$ can only be extracted to linear order in $\varepsilon$ with higher orders receiving non-perturbative contributions.  The underlying physical reason for this is due to the difference between the upper critical dimensions at YLE edge singularity ($d_c = 6$) and at the Ising critical point ($d_c = 4$).

To proceed further we will require a few well known ingredients: universal critical exponent, universal ratio $R_\chi$ and finally the scaling equation of state.  As it will become clear below,  we need to keep corrections only to the linear order in $\varepsilon$. The critical exponents  (see e.g. Ref.~\cite{BREZIN1973227}) are
\begin{align}
    \gamma&=1+\frac{1}{6}\varepsilon
+O\left(\varepsilon^2\right)\,  \\ 
\beta&= \frac{1}{2} \left( 1-\frac{1}{3}\varepsilon \right)
+O\left(\varepsilon^2\right), \\ 
\delta&=3\left(1+\frac{1}{3}\varepsilon \right)+O\left(\varepsilon^2\right)\,,
\end{align}
while the universal amplitude ratio~\cite{BREZIN1974285}
\begin{align}
    R_\chi = 1 + \frac{\varepsilon}{6} \ln \frac{27}{4} + O\left(\varepsilon^2\right)\,.
\end{align}

A convenient form  of the scaling equation of state for our purposes is given by Widom scaling (note that the mean-field magnetic equation of state~\eqref{Eq:MF} can be trivially rewritten in this form)
\begin{align}
    h = M^\delta f ( x = t M^{-1/\beta} ) 
\end{align}
which with a proper definition of the metric factors, as explained in the previous section, satisfies the following normalization conditions: 
\begin{align}
    f(0) &= 1, \\ 
    f(-1) &= 0\,. 
\end{align}
To the linear order in $\varepsilon$~\cite{1972JETPL..16..178A}, 
\begin{align}
    f(x) = 1 + x + \frac{\varepsilon}{6}\left[ 
      (x+3) \ln (x+3)
      - 3 \ln 3 + 
      x \ln \frac{4}{27} 
    \right]  + {\cal O} (\varepsilon^2)\,.
\end{align}
The presence of the non-analytic term  in the scaling equation of state is not surprising -- it encodes YLE singularity for $t>0$ and the spinodal singularities for $t<0$. In order to establish this it is convenient to consider the inverse magnetic field susceptibility $\chi^{-1} = \left. \frac{\partial h}{\partial M}\right|_t$. Its zeroes define the positions of the singularities, $x_c$. In terms of the function $f(x)$, we have 
\begin{align}
    \beta \delta f(x_c) - x_c f'(x_c) = 0\,.
\end{align}
At the leading order of $\varepsilon$-expansion, we simply obtain $x^{(0)}_c = -3$. The first correction to this approximation is already non-perturbative; it approaches zero faster then any positive fractional power of $\epsilon$ but slower than the first power. We denote this correction by  $\hat{x}(\varepsilon)$, we thus have $x_c = - 3 + \hat{x}(\varepsilon)$.  The exact expression for $\hat{x}(\varepsilon)$  is of no importance for the purpose of our discussion as we demonstrate below.  

To the first order in $\varepsilon$, 
\begin{align}
f_c = f(x_c) = -2 + \varepsilon \ln \frac32 +  \hat{x}(\varepsilon)\,.
\end{align}
Here $\hat{x}(\varepsilon)$ is the leading correction to $\varepsilon=0$ value of $f_c$. 

No we can proceed with finding $\scv_c$. For this we express $\cz$ in terms of $x$ and $f$: 
\begin{align}
    \scv_c = \frac { t_c } {h_c^{1/\beta \delta}}  = \frac{ x_c} { f_c^{1/\beta \delta} } \,.
\end{align}
The gap critical exponent $\beta \delta$ receives corrections at $\epsilon^2$ order and thus can be simply be replaced by $3/2$:  
\begin{align}
    \scv_c \approx  \frac{ -3 + \hat{x}(\varepsilon) } { (  -2 + \varepsilon \ln \frac32 +  \hat{x}(\varepsilon)  )^{2/3} } 
    \approx \scv_c^{\rm MF} \left( 1 + \frac{\varepsilon}{3} \ln \frac 32  \right)  \,.
\end{align}
Note that $\hat x(\varepsilon)$ cancels {\it exactly} at this order. We conclude that  the term of order $\varepsilon$ is the leading correction to the mean-field result: 
\begin{align}
    \scv_c = \scv_c^{\rm MF} \left( 1 + \frac{ \varepsilon}{3} \ln \frac 32  \right) + ...  
\end{align}
Here the ellipses encode non-perturbative corrections  of higher than linear in $\varepsilon$ order. 
Now for the ratio $\scv_c / R_\chi^{1/\gamma}$, using the value of $R_\chi$ computed in $\varepsilon$ expansion we get 
\begin{align}
    \frac{\scv_c}{R_\chi^{1/\gamma}} = \scv_c^{\rm MF} \left( 1 -  \frac{\varepsilon}{6} \ln 3   \right) + ...  
\end{align}
Due to the presence of the non-perturbative corrections, the conventional $\varepsilon$-expansion can  only inform us about the next-to-leading order result. 

\subsection{Two-dimensional Ising model}
\label{Sec:2d}
Onsager's solution~\cite{Onsager:1943jn} for the zero magnetic field  lattice Ising model 
complemented by the scaling theory fully determines the critical exponents of the Ising universality class in two dimensions. In the scaling limit, the correlation functions for the case of zero magnetic field was also analytically obtained \cite{Wu:1975mw}.

Further progress was achieved in the seminal work by A. Zamolodchikov~\cite{Zamolodchikov:1987zf}, see also~\cite{Zamolodchikov:1989hfa}, who found the solution of the Ising model with non-zero magnetic field at the critical temperature and directly in the scaling regime. This helped to established unknown universal quantities, including the one of interest for us -- $R_\chi$. 

Moreover, in a recent work, P.~Fonseca and A. Zamolodchikov~\cite{Fonseca:2001dc}, using the method motivated by the truncated conformal space approach~\cite{Yurov:1991my}  determined the location of the YLE singularity. Although the scaling variable of  Ref.~\cite{Fonseca:2001dc} is different from the one used in our work, the rescaling can be readily performed using known critical amplitudes. Here we describe this change of the variables and thus determine the location of the singularity of $f_G$ as defined in Section.~\ref{Sec:Scaling}. 

Up to this point we only required the magnetic equation of state and the corresponding function $f_G$. 
It fully determines the scaling behaviour including the free energy scaling equation, $f_s(t,h) = h ^ {1+\frac{1}{\delta}} f_f(\scv)$. Indeed, by taking the derivative with respect to $h$, one can establish the relation between $f_f$ and $f_G$: 
\begin{align}
    f_G(\scv) = 
    - \left( 1 + \frac{1}{\delta} \right) f_f(\scv) 
    + \frac{\scv}{\beta \delta} f'_f(\scv)\,.
\end{align}
Using the normalization property \eqref{Eq:Normalization}, we find that 
\begin{align}
f_f(0) = - \frac{ \delta } { 1 + \delta }.   \label{Eq:fZeroZ}
\end{align}
For large negative values of $\scv$, from $f_G(\scv\to-\infty) = (-\scv)^{\beta}$ we find 
\begin{align}
f_f(\scv\to-\infty) = - (-\scv)^{\beta}.   
\end{align}
These two relations are instrumental to determine the mapping from  Fonseca\--Zamo\-lodchikov's~\footnote{We preserve the notation of Ref.~\cite{Fonseca:2001dc} and use $\eta$ to denote the scaling variable only in this section; $\eta$ here should not be confused with the anomalous dimensions. } $\eta = 2\pi \tau  / h^{1/\beta\delta} = 2\pi \tau / h^{8/15}$ to $\scv$. 
The free energy in Ref.~\cite{Fonseca:2001dc} has the form 
\begin{align}
    f^{\cZ}_s (\tau,h)
     =      \pi\tau^2 \ln{ 2\pi \tau} 
     + h^{16/15} \Phi(\eta)
\end{align}
where the first term represents Onsager's singularity. 
In the limit, $\eta=0$, the scaling part of free energy 
\begin{align}
    \Phi(\eta=0) = \Phi_0  \, 
\end{align}
where the constant $\Phi_0$ is known exactly (it is an amplitude in the free energy), see Ref.~\cite{Fateev:1993av},  
\begin{align}
    \Phi_0 = 
 -{{\Gamma (1/3)\,\Gamma(1/5)\,\Gamma(7/15)}
\over{2\pi\,\Gamma(2/3)\,\Gamma(4/5)}\,\Gamma(8/15)}\,
\left(
{{4\pi^2\,\Gamma^2
(13/16)\,\Gamma(3/4)}\over {\Gamma^2
(3/16)\,\Gamma(1/4)}} \right) ^{8/15} \approx - 1.19773\,.
\end{align}
This demonstrates that $\Phi$ is not exactly equal to our conventionally defined $f_f$, as one requires an introduction of a non-trivial proportionality factor to satisfy the normalization
condition~\eqref{Eq:fZeroZ}. Consider a constant $C$, defined by $f_f(\cz) = C \Phi(\eta)$. Setting $\cz$ and $\eta$ to zero, we get  
\begin{align}
    C = - \frac{ \delta } {\Phi_0 (\delta +1) }\,.
\end{align}
Next we need to determine the relation between $\scv$ and $\eta$. For this we introduce a constant $\kappa$, $\eta = \kappa \scv$. At large negative values of the arguments from  $f_f(\scv\to   -\infty) = C \Phi (\eta \to -\infty)$ and $\Phi (\eta \to -\infty) = -\tilde G_1 (-\eta)^\beta$, where $\tilde G_1 =  - 2^{1/12} e^{-1/8} A^{3/2} \approx - 1.357838$ and $A$ is the Glaisher's constant, we establish that 
\begin{align}
    \kappa = \left( \frac{(\delta+1)\Phi_0} {\delta\,  \tilde G_1} \right)^{1/\beta}\,. 
\end{align}
This is all we need to map Fonseca-Zamolodchikov result to $z_c$. Here we  use numerical values obtained in Ref.~\cite{Xu:2022mmw}, which contain a more accurate estimate for $|\eta_{c}| = 1/\xi_c^{1/{\beta \delta}} =2.42931(7)$. It gives us  $|\scv_c| =  3.95509(12)$. Note that Fonseca-Zamolodchikov numbers would yield the value different in the fourth digit ($|\scv_c| \approx  3.9555$).  

Using well-known exact result for the universal amplitude ratio  $R_\chi = 6.778285\ldots$ (see an exhaustive list of the critical amplitude ratios for the two dimensional Ising model in Ref.~\cite{Delfino:1997ck}),  we obtain
\begin{align}
    |\scv_c|/R_\chi^{1/\gamma}  =  1.32504(4)
\end{align}
This result is plotted in Fig.~\ref{fig:YLEloc} at $d=2$.

\subsection{One-dimensional Ising model} 
\label{Sec:1d}
The one dimensional Ising model is trivially  solvable. 
Subject to some plausible assumptions one can determine the location of the YLE singularity. 

By analysing the divergence of the 
susceptibility at zero magnetic field, one can establish the existence of the critical point at $T=0$. This peculiarity of the one-dimensional Ising is not compatible with the usual definition of the scaling variable $t$. Following Fisher~\cite{Fisher:1982yc}, we consider the temperature-like variable $x$, defined as 
$x = e^{-4 J/T}$ where $J$ is the exchange integral,  and  the magnetic-like variable $h=H/T$ where $H$ is the external magnetic field.   
The free energy reads 
\begin{align}
\label{Eq:1dIsing}
f (x,h)= \ln \left[   \cosh h + \sqrt{\sinh^2 h + x }   \right] \,. 
\end{align}
The free energy in the scaling form can be obtained by performing the expansion 
\begin{align}
f_s (x,h) \approx x^{-1/2} \left( 1 + \frac{h^2}{2x}\right)  \,. 
\end{align}
Comparing this to the scaling equation of state for the free energy we conclude that, $\beta \delta = \frac{1}{2}$. Additionally, taking two derivatives with respect we obtain that $\gamma = \frac{1}{2}$. Finally, it is straightforward  to show that all critical amplitudes are equal to 1. Therefore, the scaling variable $\cz = x/h^2$ without any prefactors.  

The free energy~\eqref{Eq:1dIsing}  has an obvious square root singularity at purely imaginary values of the magnetic field 
\begin{align}
    h_c = \pm i \arcsin x ^{1/2} \,.
\end{align}
In the scaling regime, 
\begin{align}
    h_c = \pm i x ^{1/2} \, 
\end{align}
and therefore 
\begin{align}
\label{Eq:Ising1dZc} 
    |\scv_c| = \left|\frac{x}{h_c^2}\right|  = 1\,. 
\end{align}
Due to triviality of the critical amplitudes, the universal amplitude ratio $R_\chi=1$, which brings us to the final result  
\begin{align}
\label{Eq:Ising1dZetac}
    |\scv_c|/R_\chi^{1/\gamma}  =1\,.  
\end{align}

Note that there is a certain  degree of freedom in the definition of $t$, see e.g. Ref.~\cite{baxter2007exactly}. We therefore can  define the critical exponents $\gamma$  and $\beta \delta$ only up to a factor. This however does not change Eqs.~\eqref{Eq:Ising1dZc} and \eqref{Eq:Ising1dZetac}. Despite the apparent simplicity, the location of the YLE singularity in $d=1+\varepsilon$ dimensions is not known.

\section{Functional Renormalization Group}
 \label{Sec:FRG}

In order to describe the non-perturbative effects in the vicinity of the critical points, we use the FRG \cite{Wetterich:1992yh}, for related reviews see \cite{Berges:1995mw,Delamotte:2007pf,Braun:2011pp,Dupuis:2020fhh}. It is defined through the renormalization group flow equation for the scale-dependent effective action $\Gamma_k$,
\begin{align}
\label{Eq:RGFlow}
    \partial_t \Gamma_k  = 
    \frac{ 1} { 2 }
    \tilde \partial_t
    \int d^d x \left[  \ln \left(  \Gamma_k^{(2)} + R_k \right)\right]_{xx}
\end{align}
where we introduced the following convenient notation  $t = \ln {\frac{k}{\Lambda}}$
and $\tilde \partial_t = \partial_t R_k \frac{\partial} {\partial R_k}$. 
Here $R_k$ is the cut-off function. 
Throughout this paper we will utilize the (linear) Litim regulator \cite{Litim:2001up}  
\begin{align}\label{eq:reg}
    R_k(q) = \cZ_k (k^2 - q^2) \Theta(k^2 - q^2)\,,  
\end{align}
where $\cZ_k = \cZ_k (\phi_0)$ with the \emph{a priori} arbitrary expansion point $\phi_0$; we will come back to its definition in Sec.~\ref{Sect:Z}. 
Critical properties of O$(N)$ theories for real external fields have been studied with the FRG in great detail, see, e.g., \cite{Berges:1995mw, Bohr:2000gp, Litim:2001dt, Bervillier:2007rc, Braun:2007td, Braun:2009ruy, Benitez:2009xg, Stokic:2010piu, Litim:2010tt, Benitez:2011xx, Rancon_2013, Defenu:2014bea, Codello:2014yfa, Eichhorn:2016hdi, Litim:2016hlb, Juttner:2017cpr, Roscher:2018ucp, Yabunaka:2018mju, DePolsi:2020pjk}. Critical properties of the YLE have been studied in \cite{An:2016lni, Zambelli:2016cbw} and we did the first computation of $|\scv_c|$ in various critical O$(N)$ theories in \cite{Connelly:2020gwa}.

In this paper, 
we significantly improve upon our previous results in \cite{Connelly:2020gwa} by performing a full next-to-leading order computation in the derivative expansion of the FRG flow of $\Gamma_k$ with a new expansion scheme. The latter is introduced in the next section. The derivative expansion is an expansion about long-wavelength modes \cite{Morris:1994ie, Litim:2001dt} which has been demonstrated to facilitate systematic, first-principles studies of critical systems with the FRG \cite{Balog:2019rrg}. The next-to-leading order FRG flows in the derivative expansion are generated by evaluating Eq.\ \eqref{Eq:RGFlow} with the Ansatz
\begin{align}
\label{Eq:trunc}
\Gamma_k[\phi] = \int d^d x \left( U_k(\phi)  + \frac {1 } { 2 }  \cZ_k(\phi)  (\partial_i \phi)^2 \right) \,.
\end{align}
This truncation can be improved systematically by including terms of higher power in the derivatives. We note that the next-to-leading order derivative expansion is the lowest order capable of capturing a nonzero anomalous dimension. Furthermore, the regulator \eqref{eq:reg} is optimized for this truncation in the sense that it minimizes the effect of higher order corrections during the flow \cite{Litim:2000ci, Litim:2001up}. Hence, while a systematic error analysis requires the computation of higher orders in the derivative expansion as well as a study of the regulator dependence \cite{Balog:2019rrg,DePolsi:2020pjk},  the present results are obtained in a systematic way.

\subsection{Local potential approximation} 
\label{eq:lpa}

In order to set-up notations and methods, we first detail the simplest truncation for FRG calculations. The actual calculations presented in the following sections is performed using a more elaborate truncation. 

The local potential approximation (LPA) additionally assumes that $\cZ_k(\phi)\to 1$, and is therefore the leading order in the derivative expansion. Although this approximation is not bad in describing the critical properties near the Wilson-Fisher fixed point due to smallness of the anomalous dimension at $d \ge 3$, for smaller $d$ and for accessing the critical properties of the YLE singularity we expect LPA to do a poor job. For example, the leading order epsilon expansion predicts that the anomalous dimension at YLE is given by $\eta_{\rm YLE} = -(6-d)/9$~\cite{Fisher:1978pf}. That is at $d=4$,  $\eta_{\rm YLE} \approx -0.22$. The magnitude of this value is at least five times larger than $\eta_{\rm WF}$ in three dimensions.  
This indicates that, as far as the critical statics near the YLE singularity is concerned, the LPA is not a reliable approximation even in four dimensions.    
This discussion is not quantitatively rigorous, it only serves the purpose of stressing the importance of the systematic improvement of the truncation.  

That said, the {\it location} of the YLE singularity does not necessarily require the same precision in the truncation of the effective potential.  Consider $d\ge 4$. The YLE location is given by the mean-field result and known analytically, while the critical exponents at the YLE is non-trivial, e.g.\ as we mentioned above in four dimensions $\eta_{\rm YLE} \approx -0.22$. 
Our main goal is to extract the (universal) location of the YLE and for this specific purpose the LPA might be sufficient. Nonetheless, to properly account for the potential systematic effects we also consider the field-dependent wave function renormalization. 

Returning back to the truncation, we limit our consideration to homogeneous background fields. Due to translational invariance, it is  convenient to switch to momentum space. At LPA we set $\cZ_k(\phi)$ to one and substitute the truncation into the FRG equation \eqref{Eq:RGFlow}. This yields the flow of the effective potential of a homogeneous background field,
\begin{align}
    \partial_t U_k  &= 
    \frac{ 1} { 2 }
    \int \frac{d^d q}{(2\pi)^d}  \,  \tilde \partial_t \ln \left(  q^2 + \frac{\partial^2 U} {\partial^2 \phi} +  R_k(q) \right)\notag \\ 
    & = \frac{ 1} { 2 }
   \int \frac{d^d q}{(2\pi)^d}  \,  \frac{ \partial_t R_k(q)} {  q^2 + \frac{\partial^2 U} {\partial^2 \phi} +  R_k(q) } \,.
\end{align}
The regulator is especially  simple in this case
\begin{align}
    R_k(q) = (k^2 - q^2) \Theta(k^2 - q^2) 
\end{align}
with the  derivative given by 
\begin{align}
    \partial_t R_k(q) = 2 k^2 \Theta(k^2 - q^2)\,, 
\end{align}
where we note that the term related to the derivative of the $\theta$-function drops out exactly in this case. Therefore the integral in FRG equation can be computed analytically
\begin{align}
\int \frac{d^d q}{(2\pi)^d} \Theta(k^2 - q^2)
 = v_d \int_0^k  d q q^{d-1}
 = v_d \frac{k^d}{d}
\end{align}
to yield 
\begin{align}
    \partial_t U_k  &= 
    c_d  \,  
   \frac{ k^{d+2}} {  k^2 + \frac{\partial^2 U_k} {\partial^2 \phi} } \,.
\end{align}
The numerical prefactor $c_d = v_d/d = \left[ (4\pi)^{d/2} \Gamma(d/2+1)\right]^{-1} $ can be absorbed by rescaling of $U$ and $\phi$: $\tilde U = U / c_d$ and $\tilde \phi = \phi / c_d^{1/2}$. To ease the notation we drop tildes, so 
\begin{align}
    \partial_t U_k  &= 
   \frac{ k^{d+2}} {  k^2 + \frac{\partial^2 U_k} {\partial^2 \phi} } \,.
\end{align}
Finally, introducing $\rho = \phi^2/2$, the flow equation reads 
\begin{align}
    \partial_t U_k  &= 
   \frac{ k^{d+2}} {  k^2 + U_k^\prime(\rho)
   + 2 \rho U_k^{\prime\prime}(\rho)
   } \,,
\end{align}
where primes are derivatives with respect to $\rho$.

Given the initial conditions in the ultraviolet $t=0$, 
there are a few different ways to solve this partial differential equation. Here we will use a truncated Taylor expansion of the effective potential: 
\begin{align}
    U(\rho) = \sum_{i=0}^{i=N}\frac{1}{i!} a_{i}(k) (\rho - \rho_k)^i \,.  
\end{align}
where $\rho_k$ is the expansion point. Usually the expansion point is a solution of the {$k$-dependent} equation of motion for a given external field, that is $\sqrt{2\rho_k} a_1(k) = h = {\rm const}$. For the purpose of locating the YLE singularity this is not an optimal choice, as one would have to adjust the external magnetic field to tune it to the singularity. There is an additional subtle, but important drawback:
in order to perform the calculations near Wilson-Fisher fixed point, one has to initialize the flow in the broken phase in the ultraviolet (UV). Then an imaginary magnetic field, 
introduces an ambiguity in selecting the correct Riemann sheet for performing the calculations.  
Instead we fix $\rho_k$ demanding that the running mass of the critical mode is given by a fixed number ($m^2$), that is 
\begin{align}
    \label{Eq:defmin}
    a_1(k) + 2 \rho_k a_2(k) = m^2 = {\rm const}\,.
\end{align}
To locate the singularity in the symmetric phase all we need to do is to set the parameter $m^2$ to $0+$. Note that in this case the corresponding external field $h$ is not constant as a function of $k$ and assumes its genuine physical value only in the IR $k\to0$.
Thus, at this expansion point with $m^2=0$ one has
\begin{align}
    \frac{\partial U}{\partial\phi}\bigg|_{\rho_k} - h =  \frac{\partial^2 U}{\partial\phi^2}\bigg|_{\rho_k} = 0\,,
\end{align}
so that we can directly follow the flow of a critical point.

From the FRG equations, one can obtain the equations for the expansion coefficients. Using
\begin{align}
\notag  
    \partial_t U (\rho) &=
    \sum_{i=0}^{i=N}
    \left( 
    \frac{1}{i!}
    \partial_t a_{i}(k) (\rho - \rho_k)^i
-    \frac{1}{(i-1)!} a_{i}(k) (\rho - \rho_k)^{i-1} \partial_t \rho_k
     \right) \\  
     &= 
     \sum_{i=0}^{i=N}  \frac{1}{i!}
    \left( 
    \partial_t a_{i}(k) 
-     a_{i+1}(k)  \partial_t \rho_k
     \right)  (\rho - \rho_k)^{i} 
\end{align}
where we assumed that $a_{N+1}=0$. Taking the variation with respect to $\rho$ and setting it to $\rho_k$ we get the system of equations: 
\begin{align}
\label{Eq:Flowa}
\partial_t a_{i}(k) - a_{i+1}(k)  \partial_t \rho_k
 = \left[ \frac { \delta^i  \partial_t U  } { \delta \rho^i    } \right]_{\rho=\rho_k}\,, \quad i=1,\ldots,N.  \end{align} 
Explicitly, for the first and the second coefficient we obtain
\begin{align}
\label{Eq:exeq1}
\partial_t a_{1}(k) - a_{2}(k)  \partial_t \rho_k
 &= \left[\frac { \delta \partial_t U  } { \delta \rho    } \right]_{\rho=\rho_k} \,, \\ 
 \label{Eq:exeq2}
 \partial_t a_{2}(k) - a_{3}(k)  \partial_t \rho_k
 &= \left[\frac { \delta^2 \partial_t U  } { \delta \rho^2    } \right]_{\rho=\rho_k}\,.
\end{align} 
We have to complement these equations by $\partial_t \rho_k$. It can be obtained by 
differentiating Eq.~\eqref{Eq:defmin} with respect to $t$
\begin{align}
      \partial_t  a_1(k) + 2  a_2(k)  \partial_t  \rho_k + 
      2  \rho_k \partial_t  a_2(k)
      = 0 
\end{align}
which can be further simplified by using  Eqs.~\eqref{Eq:exeq1} and ~\eqref{Eq:exeq2}:
\begin{align}
\label{Eq:Flowrho}
\partial_t \rho_k 
= - 
    \frac{\left[ \frac { \delta  \partial_t U  } { \delta \rho    } \right]_{\rho=\rho_k}
     +  2  \rho_k \left[ \frac { \delta^2  \partial_t U  } { \delta \rho^2    } \right]_{\rho=\rho_k} } {3 a_{2}(k) + 2  \rho_k  a_{3}(k) }\,.
\end{align}
Note that the denominator is proportional to the third derivative of the potential with respect to the field $\partial^2 U/\partial \phi^3  = \sqrt{2\rho_k} (3 a_{2} + 2  \rho_k  a_{3}     )  $. 

Equations~\eqref{Eq:Flowa} for $i=2,\ldots, N$~\footnote{Note that the field independent part of the average potential $a_0$ is of no significance for the purpose of this paper, since we are not interested in thermodynamic properties.}, Eq.~\eqref{Eq:Flowrho} and algebraic equation~\eqref{Eq:defmin}  form a closed set of differential equations for the coefficients $a_i$ and $\rho_k$. 

The initial conditions in UV   can be chosen in multiple different ways.  The most straightforward approach is to set all coefficients $a_i(k=\Lambda)$ to zero for $i>2$ and $a_2(k=\Lambda) = {\rm const} >0$. By taking $m^2$ to zero (in practice, very small number) and varying the initial value of $\rho_{\Lambda}$ one can tune the system to the critical point $\rho_{k\to 0}=0$. Since bosonic fluctuations  drive the system towards the phase with restored symmetry, the critical point  corresponds to the initial condition with a negative  $ \rho_{\Lambda} = \rho^c_{\Lambda} <0$. The actual value of $\rho^c_{\Lambda}$  depends on $a_2(k=\Lambda)$.

Finally in order to determine the location of the YLE singularity, we simply solve the flow equations for small positive   values of $t$ and asymptotically small $m^2$. The resulting value of the magnetic field \eqref{Eq:HFlow} can then be used to extract the universal position of the singularity $\zeta_c$, see Eq.~\eqref{Eq:zetaAmpl}. One can vary $t>0$ to ensure that $\zeta_c$ is not sensitive to the value of $t$.

\subsection{Field-dependent wave-function renormalization} 
\label{Sect:Z}
In this section we go beyond LPA and consider the 
field-dependent  wave-function renormaliztion $\cZ_k(\phi)$. 
The flow for the potential can be trivially modified~\cite{Delamotte:2007pf}: 
\begin{align}\label{eq:UNLO}
    \partial_t U_k  
    & = \frac{ 1} { 2 }
   \int \frac{d^d q}{(2\pi)^d}  \,  \frac{ \partial_t R_k(q)} {  \cZ_k(\phi) q^2 + \frac{\partial^2 U} {\partial^2 \phi} +  R_k(q) } \,,
\end{align}
where we introduce the regulator 
\begin{align}
    R_k(q) = \cZ_k (k^2-q^2) \theta(k^2-q^2)
\end{align}
with $ \cZ_k =  \cZ_k (\phi_0)$; it is also convenient to operate with the renormalized field and the renormalized mass defined by 
\begin{align}
    \phi_R &= \sqrt{\cZ_k} \phi\,,\\ 
    m^2_R &= \frac{ \partial^2 U  } {\partial^2 \phi_R} = \cZ_k \frac{ \partial^2 U  } {\partial^2 \phi}\,.
\end{align}
The flow equation for the potential simplifies to 
\begin{align}
    \partial_t U_k  
    & = \frac{v_d} { 4 }
    \int_0^{k^{2}} d q^2  q^{d-2}
    \frac{ k^2 (2-\eta) + \eta q^2} 
    {k^2 + m_R^2(\phi_R) + q^2 (\cz_k(\phi_R) -1)}
\end{align}
where $\cz_k (\phi_R) = \cZ_k(\phi_R)/\cZ_k$ and $\eta = - \partial_t \ln \cZ_k$. This integral can be readily evaluated to result in 
\begin{align}
    \partial_t U_k  
    & = c_d 
    \frac{ k^{d+2} } {k^2 + m_R^2(\phi_R)} 
    \Bigg[ 
    \frac{2-\eta}{2} {}_2F_1\left( 1, \frac{d}2 \frac{d+2}2, \frac{k^2}{k^2+m_R^2(\phi_R)} (1-\cz_k(\phi_R))\right)\\ 
    &+
    \frac{1}{2} \frac{d}{d+2} \eta {}_2F_1\left( 1, \frac{d}2 , \frac{d+4}2, \frac{k^2}{k^2+m_R^2(\phi_R)} (1-\cz_k(\phi_R))\right)
    \Bigg] \,.  
\end{align}
Here ${}_2F_1$ is the hypergeometric function. 
In the limit $\eta=0$ and $\cz_k(\phi_R) =1$, we get the LPA flow equation due to ${}_2F_1\left( 1, \frac{d}2 \frac{d+2}2, 0\right) =1 $. 
In the limit, $\cz_k(\phi_R) =1$ this expression also reproduces LPA$^\prime$, i.e. (see e.g. Ref.~\cite{Delamotte:2007pf}): 
\begin{align}
    \partial_t U_k  
    & = c_d 
    \frac{ k^{d+2} } {k^2 + m_R^2(\phi_R)} 
    \Bigg(   
    1 - \frac{\eta}{d+2}
    \Bigg) \,.
\end{align}
As evident from  the truncation, for all orders of the derivative expansion 
\begin{align}
    \cZ_k(\rho)  = \lim_{p^2 \to 0 } \left. \frac{\partial \Gamma^{(2)}(p) }{ \partial p^2 } \right|_{\rho}\,.
\end{align}
The flow for $\Gamma^{(2)} (p)$ is readily obtained by variation of  Eq.~\eqref{Eq:RGFlow}: 
\begin{align}
    \partial_t \Gamma^{(2)}_k(p)
     &= 
     \frac{1}{2} \int \frac{d^d q}{(2\pi)^d} \partial_t R_k(q) G^2_k(q) \\ &\times \notag
     \left[ 
      \left( \Gamma_k^{(3)}(p,q,-p-q)  \right)^2 G_k(p+q)
     - \frac{1}{2} \Gamma_k^{(4)}(p,-p,q-q) G_k(q) 
     \right] 
\end{align}
where 
\begin{align}
    G_k(p) = \frac { 1 }{ p^2 \cZ_k(\phi) + U_k''(\phi) + R_k(p) } 
\end{align}
and the vertex  functions evaluated for the  truncation Eq.~\eqref{Eq:trunc} and the uniform field configuration
\begin{align}
    \Gamma_k^{(3)}(p,q,-p-q)  = (p^2 + q^2 + p\cdot q ) \cZ_k'(\phi) + U_k^{(3)}(\phi)
\end{align}
and 
\begin{align}
    \Gamma_k^{(4)}(p,-p,q,-q)  = (p^2+ q^2 ) \cZ_k''(\phi) + U_k^{(4)}(\phi)\,.
\end{align}
By taking the derivative with respect to $p^2$ and setting $p$ to zero we obtain
\begin{align}\notag 
    \partial_t  \cZ_k(\phi)
     &= 
    \int \frac{d^d q}{(2\pi)^d} \partial_t R_k(q) G^2_k(q) \\ &\times \notag
     \Bigg\{
     2 [q^2 \cZ_k'(\phi) + U^{(3)}(\phi)] \cZ_k'(\phi) \left[ G_k(q) + \frac{q^2}{d} \frac{\partial G_k(q)}{\partial q^2} \right] 
     \\ & +
     [q^2 \cZ_k'(\phi) + U^{(3)}(\phi)]^2 \left[  \frac{\partial G_k(q)}{\partial q^2} + 2 \frac{q^2}{d}  \frac{\partial^2 G_k(q)}{(\partial q^2)^2}\right] 
    \notag  \\ &+
     2 \frac{q^2}{d} \frac{\partial G_k(q)}{\partial q^2} \cZ_k'(\phi) U^{(3)}(\phi) - \frac12 \cZ_k''(\phi)
     \Bigg\}\,.
\end{align}
In order to obtain this expression we used the following identity:
\begin{align}
    \int d^d q (p \cdot q)^2 g(q^2) 
     = \frac{p^2}{d }  \int d^d q q^2 g(q^2)\,.
\end{align}
We now can proceed as in LPA, that is perform the expansion of $U_k(\phi)$ and $\cZ_k(\phi)$ into a power series and find flow equations 
on the expansion coefficients and the expansion point. 

There are a couple subtleties  we did not have to address in LPA.  
First, in LPA,  we fixed $m^2$ as a condition to determine the expansion point, see Eq.~\eqref{Eq:defmin}. With the field-dependent wave function renormalization, it is convenient to fix $m_R^2$. 
The corresponding flow equations are trivial to derive. Moreover, fixing $m$ may not lead to the divergence of the correlation length in IR. Indeed,  while near the YLE fixed point, the anomalous dimension is negative; therefore $\cZ_k$ tends to zero. Thus for any given $m$, at some $k$, the correlation length reaches its maximum and starts decreasing towards the IR, as the (flow-parameter dependent) correlation length $\xi_k = \frac{\sqrt{\cZ_k}}{m}$.   
Second, we fix the following normalization of $z_k$:  $\cz_k(\phi_{R,0})=1$. Separating $\cZ_k$ and $\cz_k(\phi_{R})$ created an ambiguity, this conditions is one possible way to remove it.

\subsection{Wilson-Fischer fixed point solution} 
The dimensionless flow equations are obtained by the redefinition of the field and the effective potential as follows: 
\begin{align}
    \tilde \rho & = k^{2-d} \rho_R\,, \\  
    \tilde U(\tilde \rho) & = k^{-d} U\,. 
\end{align}
The fixed point can then be found by setting the derivatives with  respect to $t$ of the expansion coefficients and $\tilde \rho_0$ to 0. This leads to a set of non-linear algebraic equations which can easily be solved numerically. For a detailed description, we refer the reader to Ref.~\cite{DePolsi:2020pjk,DePolsi:2021cmi}. The fixed point solution (for real values of $\tilde \rho_0$) describes the Wilson-Fisher fixed point. We used the set of the corresponding expansion coefficients as the initial conditions for the subsequent FRG  flow towards the IR.  The fixed point solution allows to find the anomalous dimension critical exponent directly.

\subsection{Yang-Lee edge singularity fixed point solution} \label{Sec:YLEfp}

\begin{figure}
    \centering
    \includegraphics[width=0.8\linewidth]{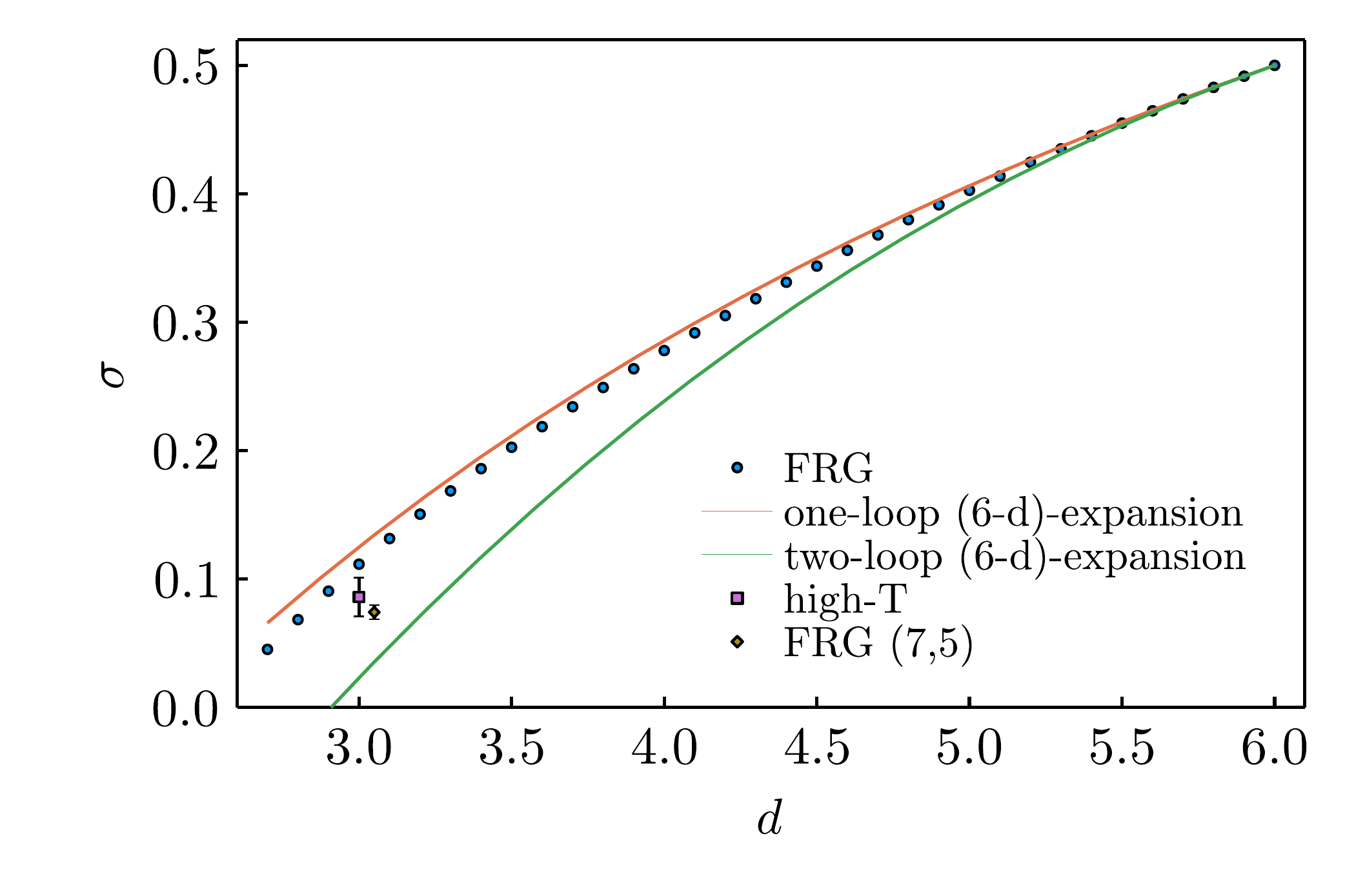}
    \caption{Dependence of the YLE critical exponent $\sigma$ on the number of dimensions. The FRG calculations are performed using the truncation $(N_{U(\phi)},N_{\cZ(\phi)}) = (8,4)$. The analytic results for the $(6-d)$-expansion are from Refs.~\cite{Fisher:1978pf,Macfarlane:1974vp}. The high temperature expansion result is taken from Ref.~\cite{Kurtze:1979zz}. 
    The FRG result obtained using the exponential regulator and the truncation  $(N_{U(\phi)},N_{\cZ(\phi)}) = (7,5)$ (note that the error bars were estimated by comparing  different truncations) is from Ref.~\cite{An:2016lni}. The data point was offset horizontally for better visibility.  For $d > 6$, $\sigma=1/2$.}
    \label{fig:sigma}
\end{figure}

For the YLE singularity fixed point it is convenient to use the expansion of the potential and the wave function renormalization  in powers of $\phi$, as was previously done in Refs.~\cite{An:2016lni,Zambelli:2016cbw}. The main remaining procedure is identical to the Wilson-Fisher fixed point, with one crucial difference: for the YLE fixed point we look for a solution for which the odd expansion coefficients are purely imaginary. As for the Wilson-Fisher fixed point, the YLE fixed point allows direct determination of the anomalous dimension. As we discussed in the introduction, at YLE there is only one truly independent critical exponent. Thus the value of YLE's $\sigma$ follows from the  scaling relation (see Eq.~\eqref{Eq:etasc}): 
\begin{align}
    \frac{1}{\sigma} = \delta_{\rm YLE} =
    \frac{d+2-\eta_{\rm YLE}}{d-2+\eta_{\rm YLE}}\,.
\end{align}
We presented the dependence of the YLE's $\sigma$ in Fig.~\ref{fig:sigma} for the truncation $(N_{U(\phi)},N_{\cZ(\phi)}) = (8,4)$. We compare with the high temperature and $6-d$-expansions~\footnote{Although currently the $\varepsilon$-expansion is known to higher number of loops, see  Ref.~\cite{Gracey:2015tta,Borinsky:2021jdb}, one would require an analysis/acceleration of its convergence, using e.g.\ a Pad\`e approximation, in order to obtain a reliable result near three dimensions}. We also included the values obtained with the FRG approach using an exponential regulator in Ref.~\cite{An:2016lni}. Despite the difference in truncation orders and the regularization scheme, our numbers are fairly close to the result of Ref.~\cite{An:2016lni}.

\subsection{Critical exponents and metric factors for $d=3$}
\label{sec:crexp}
The critical exponents can be determined by analyzing the fixed point solution. However, in order to determine the metric factors, we have to 
evolve the FRG to the IR. Since we cannot avoid computing the entire FRG evolution, we also extract  the critical exponents in the IR (as was done in Refs.~\cite{Tetradis:1993ts,Berges:1995mw}) instead of the more conventional method of diagonalizing the stability matrix at the fixed point.  

We thus now turn to solving the dynamical FRG equations. For this we need to discuss the initial conditions. 
The FRG initial conditions in the UV, $k=\Lambda$, can be chosen in multiple different ways.  The most straightforward approach is to set all coefficients $a_i(k=\Lambda)$ to zero for $i>2$ and $a_2(k=\Lambda) = {\rm const} >0$. By taking $m_R^2$ to zero (in practice, a very small number) and varying the initial value of $\rho_{\Lambda}$ one can tune the system to the critical point $\rho_{k\to 0}=0$. Since bosonic fluctuations  drive the system towards the phase with restored symmetry, the critical point  corresponds to the initial condition with a negative  $ \rho_{\Lambda} = \rho^c_{\Lambda} <0$. The actual value of $\rho^c_{\Lambda}$  depends on $a_2(k=\Lambda)$. 

To determine  critical exponents and amplitudes,  we introduce the  curvature of the potential at zero field in the UV,
\begin{align}
    r = \frac{1}{2}\frac{\partial^2 U_\Lambda(0)}  { \partial \phi^2}  = 
\frac{m_R^2 - 3a_2(\Lambda) \rho_\Lambda}{2}\,,
\end{align}
and its critical value 
\begin{align}
    r_c =  
- \frac{ 3} {2} a_2(\Lambda) \rho^c_\Lambda. \end{align}
The deviation from the critical temperature is proportional to the difference  $t = r - r_c$. 

In order to find the critical exponent  $\delta$, we adjust the parameter $m_R^2$ such that the system stays on the critical isotherm $t=0$, i.e. $r = r_c$, while varying $\rho_\Lambda$. Then from the relation 
\begin{align}
\label{Eq:HFlow}
 H = \lim_{k\to0} \frac{\partial U}{\partial \phi }  = 
 \lim_{k\to0} \sqrt{\cZ(k)} \frac{\partial U}{\partial \phi_R } = 
 \lim_{k\to0} \sqrt{ 2 \rho_{k} \cZ(k)} a_1(k) 
\end{align}
we obtain the value of the magnetic field, while the magnetization is given by $M = \sqrt{2\rho_0/\cZ(0)}$. The power law dependence of $M$ on $h$ defines the critical exponent $\delta$ and the amplitude $B_c$, see Eq.~\eqref{Eq:M_tc}.  

In order to find the critical exponent  $\gamma$, we consider positive values of $t>0$ (symmetric phase). While the parameters $m_R^2$ and $\rho_\Lambda$ are fixed in a such a way as to yield zero magnetization in the IR $\rho_0 =0$. In practice it is convenient to first fix $m_R^2$ and then by adjusting $\rho_\Lambda$ ensure that $\rho_0 =0$. This procedure is to be repeated for a set of $m_R^2$. 
The magnetic field susceptibility is then given by $m_R^2$, while $t$ is defined by the initial conditions. This complemented by  
Eq.~\eqref{Eq:Chi} leads to determination of the critical exponent $\gamma$ and  and the amplitude $C_+$. 
 
In performing the actual fits, we take into account the leading corrections to scaling (see e.g. Ref.~\cite{Amit:1984ms}), which can be encoded by the critical exponents $\theta_t = \omega\nu$  and $\theta_h$: 
\begin{align}
    \chi &= C_+ t^{-\gamma} \left( 1 + a_t 
    t^{\theta_t}  
    \right), \\ 
    M &= B_c H^{1/\delta}\left( 1 + a_h 
    H^{\theta_h}  
    \right)\,.
\end{align}
We treat $C^+$, $\gamma$, $\theta_{t,h}$ and $a_{t,h}$ as independent fitting parameters. The exponents are however not independent, indeed, the scaling relations predict that 
\begin{align}
\label{Eq:thetas}
    \theta_t  = \Delta \cdot \theta_h 
\end{align}
where the gap critical exponent 
\begin{align}
 \Delta = \beta \delta = \frac{\gamma \delta}{\delta-1}\,.
\end{align}
We use the relation \eqref{Eq:thetas} to cross-check the sensibility of our fits. 

For the truncation $(N_{U(\phi)},N_{\cZ(\phi)}) = (8,4)$  we obtain 
$\theta_t/\theta_h = 1.550358$ exclusively from the corrections to scaling, while from the leading order scaling we get $\gamma \delta /(\delta -1)= 1.550347$. That shows that the precision of our numerical calculations is sufficient to capture the corrections to leading order scaling adequately.  In order to cross-check the hyper-scaling relations, we also compare the anomalous dimension evaluated at the fixed point with the one obtained from the scaling relation \eqref{Eq:etasc}. In three dimensions with the same truncation as above, the difference between the anomalous dimensions obtained in these two ways is of order $10^{-12}$. 
Although this does not represent the systematical error of our calculations, 
it demonstrates the internal consistency of our numerical method. 

We list the leading order critical exponents and the metric factors in Table~\ref{tab:crit} for  $d=3$ and a few representative truncations.
The numerical values for the critical exponents are in agreement with the corresponding results at next-to-leading order derivative expansion in Refs.~\cite{Balog:2019rrg, DePolsi:2020pjk}. As shown in these works, the precision can be  systematically improved through the inclusion of higher orders in the derivative expansion leading to an excellent agreement with the results of conformal bootstrap \cite{Kos:2016ysd}.

\begin{table}[]
    \centering
    \begin{tabular}{c|c|c|c|c}
         Order $(N_{U(\phi)}, N_{\cZ(\phi)})$&  $\delta$ 
         & $\gamma$  & $B_c$ & $C_+$ 
         \\
         \hline 
         (10,4) & 4.710 & 1.221 & 3.069 & 1.104 \\ 
         \hline 
         (8,4) &  4.714 & 1.221 & 3.041 & 1.104 \\ 
         \hline 
         (8,6) &  4.714 & 1.220 & 3.039  & 1.106
    \end{tabular}
    \caption{Critical exponents and amplitudes for three different truncation schemes at $d=3$. In numerical values, we present three significant digits in order to demonstrate the variation of the critical exponents and amplitudes  for different truncation orders.
    See Refs.~\cite{Balog:2019rrg, DePolsi:2020pjk} for state of the art high precision calculations of the critical  exponents with the FRG. 
    }
    \label{tab:crit}
\end{table}

\subsection{Location of  Yang-Lee edge singularity for $2.7\le d \le4$}
As we alluded to before, in order to extract the YLE location, we compute in the symmetric phase  $t>0$ and at  $m_R = 0+$. We are interested in defining the universal location, thus, in order to minimize non-universal contributions we consider rather small $t$. 

One complication arises. On the one hand, the most suitable truncation for working near the YLE singularity is in terms of $\phi$. On the other hand, the calculations near the Wilson-Fisher fixed point and the crossing of $\rho_0$ from positive (real $\phi_0$) to negative (purely imaginary $\phi_0$) values are more straightforward to formulate in  terms of $\rho$. We solve this dilemma by performing calculations within an expansion in  $\rho$ until the scale-dependent anomalous dimension reaches small negative values. At this point we perform the matching to the expansion 
\begin{align}
    U(\phi) = \sum_{i=0}^{i=2N}\frac{1}{i!} b_{i}(k) (\phi - \phi_k)^i \,,  
\end{align}
and similarly for $\cZ$. In order to perform the transition from $\rho$ to $\phi$, one has to double the number of expansion coefficients. To illustrate the procedure, lets consider some 
$k^*$ at which $\rho(k^*) <0$, then we choose a positive root $\phi_{k^*} = i \sqrt{-\rho_{k^*}}$ for definiteness. The expansion coefficients are then easily found, for example 
\begin{align}
    b_1(k^*) &= \phi_{k^*} \, a_1(k^*)\,, \\ 
    b_2(k^*) &=  a_1(k^*) + 2 \rho(k^*) a_2(k^*)\,, \\ 
    b_3(k^*) &=  \phi_{k^*} \left[ a_2(k^*) + 2 \rho(k^*) a_3(k^*) \right]\,,
\end{align}
 etc. 
Although the flow equations decouple from $b_1(k)$, we still
need to solve for it in order to determine the value of the magnetic field in the IR.

\begin{figure}
    \centering
    \includegraphics[width=0.8\linewidth]{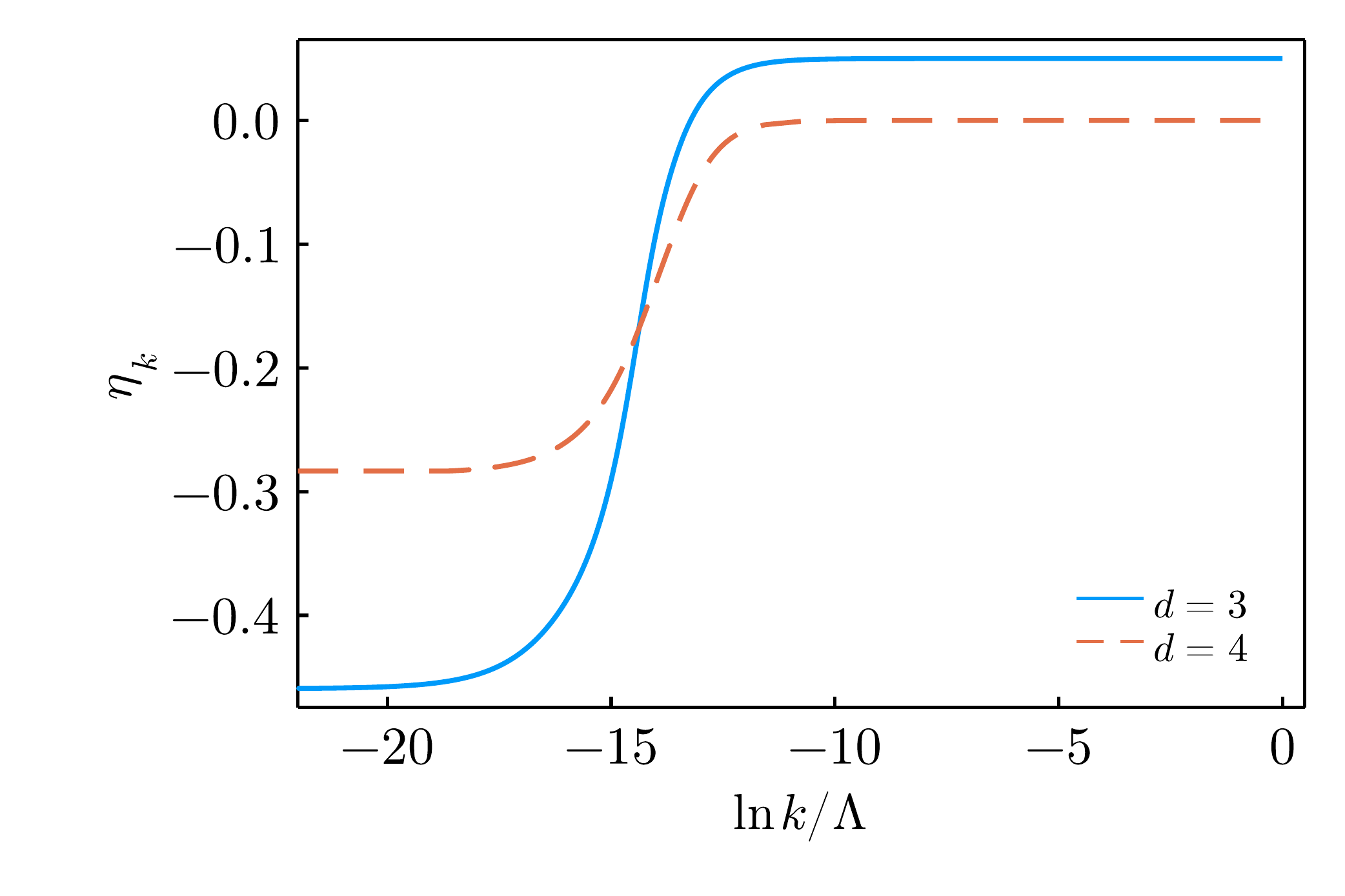}
    \caption{Evolution of the anomalous dimension as a function of the flow ``time''. 
    The UV ($k=\Lambda$) initial condition is chosen to be very close to the Wilson Fisher point. Thus $\eta_\Lambda \approx \eta_{\rm WF}$. In the IR the system approaches the YLE fixed point, $\eta_{k \to 0} \approx \eta_{\rm YLE}$. The dashed curve at $d=4$ demonstrates the transition from a trivial Gaussian fixed point to the YLE fixed point. The figure demonstrates the crossover between the two fixed points. 
    } 
    \label{fig:YLEevo}
\end{figure}

We checked that, as long as we keep it small,  the exact value of the anomalous dimension at which we perform switching $\eta(k^*)<0$ does not affect the values of the coupling and the location of the singularity in the IR.

We demonstrate the dependence of the scale-dependent anomalous dimension on $k$ for $d=3$ and 4 in Fig.~\ref{fig:YLEevo}. The initial conditions are chosen in the symmetric phase and very close to the fixed point. As the figure demonstrates, the system lingers near the Wilson-Fisher  for a significant range of $k$, then it smoothly transitions to the YLE fixed point. Since we chose a very small value of $m_R^2 =  10^{-40}$ in this calculation, the  deviation from the YLE fixed point will only be visible at $\ln k /\Lambda \lesssim  -90$.  
We remind the reader that in four dimensions, the Wilson-Fisher point can be described using the mean-field approximation with the characteristic  zero anomalous dimensions. Our calculations demonstrate the transition from this trivial case to the YLE fixed point with non-zero $\eta$. This reflects the fact that the upper critical dimension for the YLE fixed point is six, and below this dimension critical fluctuations (as reflected by the anomalous dimensions) play an important role.
The transition from Wilson-Fisher to the YLE fixed point is explicitly computed for the first time in this paper.  

The fact that the FRG evolution brings us to the fixed point with a negative $\eta$ gives us confidence that we indeed extract the location of the YLE singularity.  
Note that the same values of $\eta$ can be obtained by solving algebraic equations at the YLE fixed point, see Sec.~\ref{Sec:YLEfp}.

\begin{figure}[ht]
    \centering
    \includegraphics[width=0.8\linewidth]{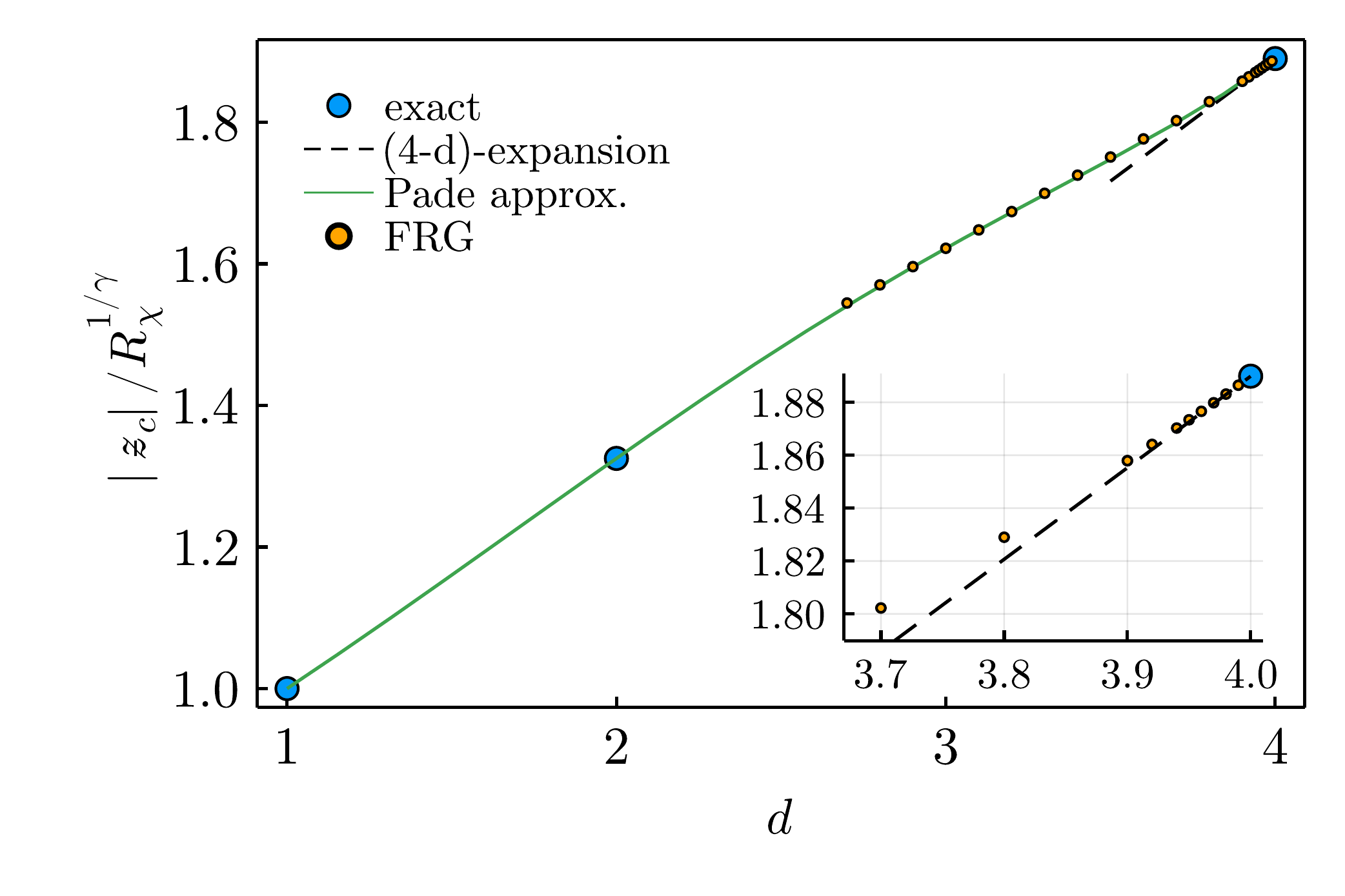}
    \caption{The universal location of the YLE singularity as a function of the number of dimensions $d$. See Sec.~\ref{Sec:Analytic} for (semi-)exact results obtained at $d=1, 2$, and $d\ge 4$. See Sec.~\ref{Sec:Epsilon} for the leading order $(4-d)$-expansion. 
    The FRG points are obtained with the truncation $(N_{U(\phi), N_{\cZ(\phi)}}) = (8,4)$.
    Variations in $N_{U(\phi)}$ by $\pm 2$ produces deviations consistent with the symbol size at any computed $d$. The variation in $N_{\cZ(\phi)}$ by $\pm2$ at $d=3$ is also consistent with the symbol size. The solid curve displays a plausible interpolation of the location of the YLE singularity for any $d>1$. It was obtained using  Pad\'e approximation with  $d=1,2,4$, the leading order $(4-d)$-expansion, 
    and the FRG results at $d=3$ as an input. The inset figure magnifies the region near $d=4$ to demonstrate that the FRG results coincides with the $(4-d)$-expansion close to $d=4$.  
    }
    \label{fig:YLEloc}
\end{figure}

By evaluating the required combination, see Eq.~\eqref{Eq:zetaAmpl}, we found the YLE singularity at various $d>2.7$ and compared them to the analytical results in Fig.~\ref{fig:YLEloc}. For smaller $d$, solving FRG equations  becomes  computationally more expensive. At the same time, the higher order derivative expansion coefficients might also start to play a dominant role.  In   Fig.~\ref{fig:YLEloc}, we also show the inset demonstrating that, our results well reproduce the $\varepsilon$-expansion near 4 dimensions. To provide a viable dependence of the YLE singularity location on the number of dimensions, we also plotted the Pad\`e approximation
using  $d=1,2,4$, the leading order $(4-d)$-expansion,    and the FRG results at $d=3$ as an input. We also require for the Pad\'e approximation to have neither zeroes nor poles for $1\le d \le 4$. This fixes the minimal order $[1/3]$ for the Pad\'e approximant~\footnote{The explicit form is $P[1/3] \approx \frac{0.724911 - 0.0957697 d} {1 - 0.459701 d + 0.0972128 d^2 - 0.00837074 d^3} $}. 

Table~\ref{tab:yleunc} summarizes the YLE singularity location for different truncation orders in three dimensions. The dependence on the truncation allows us to establish the uncertainty of its determination. We get $|\scv_c|/R_\chi^{1/\gamma} =  1.621(3)$. This uncertainty is smaller then the size of the corresponding marker in Fig.~\ref{fig:YLEloc}. We note that it represents the uncertainty in our computation of the full next-to-leading order derivative expansion, which entails a fully field-dependent wave function renormalization, $\cZ_k(\phi)$, and effective potential $U_k(\phi)$, see Eq.~\eqref{Eq:trunc}. As discussed in Sec.~\ref{Sec:FRG}, a systematic error analysis would require computation of higher orders in the derivative expansion and an estimate of the regulator dependence \cite{Balog:2019rrg, DePolsi:2020pjk}.

\begin{table}[t]
    \centering
    \begin{tabular}{c|c}
         Order $(N_{U(\phi)}, N_{\cZ(\phi)})$&  \quad YLE Location $|\scv_c|/R_\chi^{1/\gamma}$ \\
         \hline 
         (10,4) & 1.6228 \\ 
         \hline 
         (8,4) &  1.6219 \\ 
         \hline 
         (8,6) &  1.6246
    \end{tabular}
    \caption{YLE singularity locations computed from three different truncation schemes at $d=3$. }
    \label{tab:yleunc}
\end{table}

\section{Conclusions}
In this paper we presented analytical results and numerical data obtained within the systematic framework of the FRG with a next-to-leading order derivative expansion, establishing the location of the Yang-Lee edge singularity for the Ising universality class.

Our results span the entire domain of the non-trivial critical behaviour $1\le d \le 4$. At $d=4$, the location of the Yang-Lee edge singularity was known; it can be easily extracted using the mean-field approximation. 
Near four dimensions, $d=4-\varepsilon$, we performed analysis of the equation of state in $\varepsilon$-expansions and were able to determine the first non-trivial perturbative contribution. 
At $d=2$, we relied on the results of Fonseca, Xu and Zamolodchikov \cite{Fonseca:2001dc,Xu:2022mmw}.
We mapped their result onto the more commonly used scaling variable also used here.  At $d=1$, we performed a trivial analysis to extract the location of the singularity. There are no known analytical methods to address the problem in three dimensions. We thus used a numerical first-principles approach -- the FRG. We analyzed the precision of our truncation scheme and found that the location of the Yang Lee edge singularity is $|\scv_c|/R_\chi^{1/\gamma} = 1.621(3)$ in three dimensions. We also performed calculations in fractional dimensions  in order to demonstrate that they reproduce the $\varepsilon$ expansion near four dimensions.  

Additionally, by performing direct numerical calculations,   we explicitly showed the transition between Wilson-Fisher fixed point (positive anomalous dimension) and the YLE fixed point (negative anomalous dimension). Our result demonstrates that one can also study a corresponding {\it crossover} behaviour between these two fixed points.    

As a by product of our calculations, we extracted the anomalous dimensions at the YLE. Using the Litim regulator we obtained the numbers consistent with the one extracted in Ref.~\cite{An:2016lni} using the   exponential regulator. We note that we did not have any issues location the Yang-Lee edge fixed point using the Litim regulator at any $d>2.7$ 
in contrast to Ref.~\cite{Zambelli:2016cbw}.

The application of our results to QCD is two fold. Both are related to the underlying $\cZ(2)$ universality class.  First, the universal location of the YLE singularity might facilitate the discovery of the QCD critical point at finite real baryon chemical potential, which lies in the same universality class. Second, one can study the property of the Roberge-Weiss critical point~\cite{Roberge:1986mm} at purely imaginary baryon chemical potential. We refer the reader to the recent model and lattice QCD studies in Ref.~\cite{Basar:2021hdf,Basar:2021gyi,Singh:2021pog,Dimopoulos:2021vrk,Attanasio:2021tio,Mukherjee:2019eou}.

\section{ Acknowledgement  }
We thank G.~Basar,  A.~Connelly,  A.~Kemper, J.~Pawlowski,  and C.~Schmidt  for useful  discussions, {and G.~Johnson for the collaboration on related subjects}. 
We are  grateful
to  B.~Friman and, especially, S.~Mukherjee for stimulating discussions leading to this work.  
We acknowledge the computing resources provided on Henry2, a high-performance computing cluster operated by North Carolina State University, and support by  the U.S. Department of Energy, Office of Science,
Office of Nuclear Physics through the Contract No. DE-SC0020081.

\appendix

\bibliography{YLEIsing}

\begin{thebibliography}{10}
\expandafter\ifx\csname url\endcsname\relax
  \def\url#1{\texttt{#1}}\fi
\expandafter\ifx\csname urlprefix\endcsname\relax\def\urlprefix{URL }\fi
\expandafter\ifx\csname href\endcsname\relax
  \def\href#1#2{#2} \def\path#1{#1}\fi

\bibitem{PhysRev.87.404}
C.~N. Yang, T.~D. Lee,
  \href{https://link.aps.org/doi/10.1103/PhysRev.87.404}{Statistical theory of
  equations of state and phase transitions. i. theory of condensation}, Phys.
  Rev. 87 (1952) 404--409.
\newblock \href {https://doi.org/10.1103/PhysRev.87.404}
  {\path{doi:10.1103/PhysRev.87.404}}.
\newline\urlprefix\url{https://link.aps.org/doi/10.1103/PhysRev.87.404}

\bibitem{PhysRev.87.410}
T.~D. Lee, C.~N. Yang,
  \href{https://link.aps.org/doi/10.1103/PhysRev.87.410}{Statistical theory of
  equations of state and phase transitions. ii. lattice gas and ising model},
  Phys. Rev. 87 (1952) 410--419.
\newblock \href {https://doi.org/10.1103/PhysRev.87.410}
  {\path{doi:10.1103/PhysRev.87.410}}.
\newline\urlprefix\url{https://link.aps.org/doi/10.1103/PhysRev.87.410}

\bibitem{PhysRevLett.27.1439}
P.~J. Kortman, R.~B. Griffiths,
  \href{https://link.aps.org/doi/10.1103/PhysRevLett.27.1439}{Density of zeros
  on the lee-yang circle for two ising ferromagnets}, Phys. Rev. Lett. 27
  (1971) 1439--1442.
\newblock \href {https://doi.org/10.1103/PhysRevLett.27.1439}
  {\path{doi:10.1103/PhysRevLett.27.1439}}.
\newline\urlprefix\url{https://link.aps.org/doi/10.1103/PhysRevLett.27.1439}

\bibitem{Fisher:1978pf}
M.~E. Fisher, {Yang-Lee Edge Singularity and phi**3 Field Theory}, Phys. Rev.
  Lett. 40 (1978) 1610--1613.
\newblock \href {https://doi.org/10.1103/PhysRevLett.40.1610}
  {\path{doi:10.1103/PhysRevLett.40.1610}}.

\bibitem{bena2005statistical}
I.~Bena, M.~Droz, A.~Lipowski, Statistical mechanics of equilibrium and
  nonequilibrium phase transitions: the yang--lee formalism, International
  Journal of Modern Physics B 19~(29) (2005) 4269--4329.

\bibitem{Itzykson:1983gb}
C.~Itzykson, R.~B. Pearson, J.~B. Zuber, {Distribution of Zeros in Ising and
  Gauge Models}, Nucl. Phys. B220 (1983) 415--433.
\newblock \href {https://doi.org/10.1016/0550-3213(83)90499-6}
  {\path{doi:10.1016/0550-3213(83)90499-6}}.

\bibitem{Fisher:1982yc}
M.~E. Fisher, {SCALING, UNIVERSALITY AND RENORMALIZATION GROUP THEORY} (1982).

\bibitem{Gliozzi:2014jsa}
F.~Gliozzi, A.~Rago, {Critical exponents of the 3d Ising and related models
  from Conformal Bootstrap}, JHEP 10 (2014) 042.
\newblock \href {http://arxiv.org/abs/1403.6003} {\path{arXiv:1403.6003}},
  \href {https://doi.org/10.1007/JHEP10(2014)042}
  {\path{doi:10.1007/JHEP10(2014)042}}.

\bibitem{Gracey:2015tta}
J.~A. Gracey, {Four loop renormalization of $\phi^3$ theory in six dimensions},
  Phys. Rev. D 92~(2) (2015) 025012.
\newblock \href {http://arxiv.org/abs/1506.03357} {\path{arXiv:1506.03357}},
  \href {https://doi.org/10.1103/PhysRevD.92.025012}
  {\path{doi:10.1103/PhysRevD.92.025012}}.

\bibitem{Borinsky:2021jdb}
M.~Borinsky, J.~A. Gracey, M.~V. Kompaniets, O.~Schnetz, {Five-loop
  renormalization of \ensuremath{\phi}3 theory with applications to the
  Lee-Yang edge singularity and percolation theory}, Phys. Rev. D 103~(11)
  (2021) 116024.
\newblock \href {http://arxiv.org/abs/2103.16224} {\path{arXiv:2103.16224}},
  \href {https://doi.org/10.1103/PhysRevD.103.116024}
  {\path{doi:10.1103/PhysRevD.103.116024}}.

\bibitem{An:2016lni}
X.~An, D.~Mesterh\'azy, M.~A. Stephanov, {Functional renormalization group
  approach to the Yang-Lee edge singularity}, JHEP 07 (2016) 041.
\newblock \href {http://arxiv.org/abs/1605.06039} {\path{arXiv:1605.06039}},
  \href {https://doi.org/10.1007/JHEP07(2016)041}
  {\path{doi:10.1007/JHEP07(2016)041}}.

\bibitem{Zambelli:2016cbw}
L.~Zambelli, O.~Zanusso, {Lee-Yang model from the functional renormalization
  group}, Phys. Rev. D 95~(8) (2017) 085001.
\newblock \href {http://arxiv.org/abs/1612.08739} {\path{arXiv:1612.08739}},
  \href {https://doi.org/10.1103/PhysRevD.95.085001}
  {\path{doi:10.1103/PhysRevD.95.085001}}.

\bibitem{Cardy:1996xt}
J.~L. Cardy, {Scaling and renormalization in statistical physics}, 1996.

\bibitem{Amit:1984ms}
D.~J. Amit, {FIELD THEORY, THE RENORMALIZATION GROUP, AND CRITICAL PHENOMENA},
  1984.

\bibitem{henrici1991applied}
P.~Henrici, \href{https://books.google.com/books?id=KPpQAAAAMAAJ}{Applied and
  Computational Complex Analysis, Volume 2: Special Functions, Integral
  Transforms, Asymptotics, Continued Fractions}, Applied and Computational
  Complex Analysis, Wiley, 1991.
\newline\urlprefix\url{https://books.google.com/books?id=KPpQAAAAMAAJ}

\bibitem{Stephanov:2006dn}
M.~A. Stephanov, {QCD critical point and complex chemical potential
  singularities}, Phys. Rev. D 73 (2006) 094508.
\newblock \href {http://arxiv.org/abs/hep-lat/0603014}
  {\path{arXiv:hep-lat/0603014}}, \href
  {https://doi.org/10.1103/PhysRevD.73.094508}
  {\path{doi:10.1103/PhysRevD.73.094508}}.

\bibitem{Mukherjee:2019eou}
S.~Mukherjee, V.~Skokov, {Universality driven analytic structure of the QCD
  crossover: radius of convergence in the baryon chemical potential}, Phys.
  Rev. D 103~(7) (2021) L071501.
\newblock \href {http://arxiv.org/abs/1909.04639} {\path{arXiv:1909.04639}},
  \href {https://doi.org/10.1103/PhysRevD.103.L071501}
  {\path{doi:10.1103/PhysRevD.103.L071501}}.

\bibitem{Basar:2021gyi}
G.~Basar, G.~Dunne, Z.~Yin, {Uniformizing Lee-Yang Singularities} (12 2021).
\newblock \href {http://arxiv.org/abs/2112.14269} {\path{arXiv:2112.14269}}.

\bibitem{Basar:2021hdf}
G.~Basar, {Universality, Lee-Yang Singularities, and Series Expansions}, Phys.
  Rev. Lett. 127~(17) (2021) 171603.
\newblock \href {http://arxiv.org/abs/2105.08080} {\path{arXiv:2105.08080}},
  \href {https://doi.org/10.1103/PhysRevLett.127.171603}
  {\path{doi:10.1103/PhysRevLett.127.171603}}.

\bibitem{Giordano:2019gev}
M.~Giordano, K.~Kapas, S.~D. Katz, D.~Nogradi, A.~Pasztor, {Radius of
  convergence in lattice QCD at finite $\mu_B$ with rooted staggered fermions},
  Phys. Rev. D 101~(7) (2020) 074511.
\newblock \href {http://arxiv.org/abs/1911.00043} {\path{arXiv:1911.00043}},
  \href {https://doi.org/10.1103/PhysRevD.101.074511}
  {\path{doi:10.1103/PhysRevD.101.074511}}.

\bibitem{Attanasio:2021tio}
F.~Attanasio, M.~Bauer, L.~Kades, J.~M. Pawlowski, {Searching for Yang-Lee
  zeros in O($N$) models}, in: {38th International Symposium on Lattice Field
  Theory}, 2021.
\newblock \href {http://arxiv.org/abs/2111.12645} {\path{arXiv:2111.12645}}.

\bibitem{Nicotra:2021ijp}
G.~Nicotra, P.~Dimopoulos, L.~Dini, F.~Di~Renzo, J.~Goswami, C.~Schmidt,
  S.~Singh, K.~Zambello, F.~Ziesche, {Lee-Yang edge singularities in 2+1 flavor
  QCD with imaginary chemical potential}, in: {38th International Symposium on
  Lattice Field Theory}, 2021.
\newblock \href {http://arxiv.org/abs/2111.05630} {\path{arXiv:2111.05630}}.

\bibitem{Dimopoulos:2021vrk}
P.~Dimopoulos, L.~Dini, F.~Di~Renzo, J.~Goswami, G.~Nicotra, C.~Schmidt,
  S.~Singh, K.~Zambello, F.~Ziesch\'e, {Contribution to understanding the phase
  structure of strong interaction matter: Lee-Yang edge singularities from
  lattice QCD}, Phys. Rev. D 105~(3) (2022) 034513.
\newblock \href {http://arxiv.org/abs/2110.15933} {\path{arXiv:2110.15933}},
  \href {https://doi.org/10.1103/PhysRevD.105.034513}
  {\path{doi:10.1103/PhysRevD.105.034513}}.

\bibitem{Ejiri:2005ts}
S.~Ejiri, {Lee-Yang zero analysis for the study of QCD phase structure}, Phys.
  Rev. D 73 (2006) 054502.
\newblock \href {http://arxiv.org/abs/hep-lat/0506023}
  {\path{arXiv:hep-lat/0506023}}, \href
  {https://doi.org/10.1103/PhysRevD.73.054502}
  {\path{doi:10.1103/PhysRevD.73.054502}}.

\bibitem{Wei:2016oyn}
B.-B. Wei, {Probing Yang\textendash{}Lee edge singularity by central spin
  decoherence}, New J. Phys. 19~(8) (2017) 083009.
\newblock \href {http://arxiv.org/abs/1611.08074} {\path{arXiv:1611.08074}},
  \href {https://doi.org/10.1088/1367-2630/aa77d6}
  {\path{doi:10.1088/1367-2630/aa77d6}}.

\bibitem{francis2021manybody}
A.~Francis, D.~Zhu, C.~H. Alderete, S.~Johri, X.~Xiao, J.~K. Freericks,
  C.~Monroe, N.~M. Linke, A.~F. Kemper,
  \href{https://www.science.org/doi/abs/10.1126/sciadv.abf2447}{Many-body
  thermodynamics on quantum computers via partition function zeros}, Science
  Advances 7~(34) (2021) eabf2447.
\newblock \href
  {http://arxiv.org/abs/https://www.science.org/doi/pdf/10.1126/sciadv.abf2447}
  {\path{arXiv:https://www.science.org/doi/pdf/10.1126/sciadv.abf2447}}, \href
  {https://doi.org/10.1126/sciadv.abf2447} {\path{doi:10.1126/sciadv.abf2447}}.
\newline\urlprefix\url{https://www.science.org/doi/abs/10.1126/sciadv.abf2447}

\bibitem{Connelly:2020gwa}
A.~Connelly, G.~Johnson, F.~Rennecke, V.~Skokov, {Universal Location of the
  Yang-Lee Edge Singularity in $O(N)$ Theories}, Phys. Rev. Lett. 125~(19)
  (2020) 191602.
\newblock \href {http://arxiv.org/abs/2006.12541} {\path{arXiv:2006.12541}},
  \href {https://doi.org/10.1103/PhysRevLett.125.191602}
  {\path{doi:10.1103/PhysRevLett.125.191602}}.

\bibitem{Morris:1994ie}
T.~R. Morris, {Derivative expansion of the exact renormalization group}, Phys.
  Lett. B 329 (1994) 241--248.
\newblock \href {http://arxiv.org/abs/hep-ph/9403340}
  {\path{arXiv:hep-ph/9403340}}, \href
  {https://doi.org/10.1016/0370-2693(94)90767-6}
  {\path{doi:10.1016/0370-2693(94)90767-6}}.

\bibitem{Litim:2001dt}
D.~F. Litim, {Derivative expansion and renormalization group flows}, JHEP 11
  (2001) 059.
\newblock \href {http://arxiv.org/abs/hep-th/0111159}
  {\path{arXiv:hep-th/0111159}}, \href
  {https://doi.org/10.1088/1126-6708/2001/11/059}
  {\path{doi:10.1088/1126-6708/2001/11/059}}.

\bibitem{Balog:2019rrg}
I.~Balog, H.~Chat\'e, B.~Delamotte, M.~Marohnic, N.~Wschebor, {Convergence of
  Nonperturbative Approximations to the Renormalization Group}, Phys. Rev.
  Lett. 123~(24) (2019) 240604.
\newblock \href {http://arxiv.org/abs/1907.01829} {\path{arXiv:1907.01829}},
  \href {https://doi.org/10.1103/PhysRevLett.123.240604}
  {\path{doi:10.1103/PhysRevLett.123.240604}}.

\bibitem{DePolsi:2020pjk}
G.~De~Polsi, I.~Balog, M.~Tissier, N.~Wschebor, {Precision calculation of
  critical exponents in the $O(N)$ universality classes with the
  nonperturbative renormalization group}, Phys. Rev. E 101~(4) (2020) 042113.
\newblock \href {http://arxiv.org/abs/2001.07525} {\path{arXiv:2001.07525}},
  \href {https://doi.org/10.1103/PhysRevE.101.042113}
  {\path{doi:10.1103/PhysRevE.101.042113}}.

\bibitem{Zinn-Justin:2002ecy}
J.~Zinn-Justin, {Quantum field theory and critical phenomena}, Int. Ser.
  Monogr. Phys. 113 (2002) 1--1054.

\bibitem{Vasilev:2004yr}
A.~N. Vasilev, {The field theoretic renormalization group in critical behavior
  theory and stochastic dynamics}, 2004.

\bibitem{Caselle:2000nn}
M.~Caselle, M.~Hasenbusch, A.~Pelissetto, E.~Vicari, {The critical equation of
  state of the 2-D ising model}, J. Phys. A 34 (2001) 2923--2948.
\newblock \href {http://arxiv.org/abs/cond-mat/0011305}
  {\path{arXiv:cond-mat/0011305}}, \href
  {https://doi.org/10.1088/0305-4470/34/14/302}
  {\path{doi:10.1088/0305-4470/34/14/302}}.

\bibitem{Fonseca:2001dc}
P.~Fonseca, A.~Zamolodchikov, {Ising field theory in a magnetic field: Analytic
  properties of the free energy} (12 2001).
\newblock \href {http://arxiv.org/abs/hep-th/0112167}
  {\path{arXiv:hep-th/0112167}}.

\bibitem{PhysRevE.65.066127}
M.~Campostrini, A.~Pelissetto, P.~Rossi, E.~Vicari,
  \href{https://link.aps.org/doi/10.1103/PhysRevE.65.066127}{25th-order
  high-temperature expansion results for three-dimensional ising-like systems
  on the simple-cubic lattice}, Phys. Rev. E 65 (2002) 066127.
\newblock \href {https://doi.org/10.1103/PhysRevE.65.066127}
  {\path{doi:10.1103/PhysRevE.65.066127}}.
\newline\urlprefix\url{https://link.aps.org/doi/10.1103/PhysRevE.65.066127}

\bibitem{An:2017brc}
X.~An, D.~Mesterh\'azy, M.~A. Stephanov, {On spinodal points and Lee-Yang edge
  singularities}, J. Stat. Mech. 1803~(3) (2018) 033207.
\newblock \href {http://arxiv.org/abs/1707.06447} {\path{arXiv:1707.06447}},
  \href {https://doi.org/10.1088/1742-5468/aaac4a}
  {\path{doi:10.1088/1742-5468/aaac4a}}.

\bibitem{BREZIN1973227}
E.~Brezin, J.~{Le Guillou}, J.~Zinn-Justin, B.~Nickel,
  \href{https://www.sciencedirect.com/science/article/pii/0375960173908943}{Higher
  order contributions to critical exponents}, Physics Letters A 44~(3) (1973)
  227--228.
\newblock \href {https://doi.org/https://doi.org/10.1016/0375-9601(73)90894-3}
  {\path{doi:https://doi.org/10.1016/0375-9601(73)90894-3}}.
\newline\urlprefix\url{https://www.sciencedirect.com/science/article/pii/0375960173908943}

\bibitem{BREZIN1974285}
E.~Brezin, J.-C. {Le Guillou}, J.~Zinn-Justin,
  \href{https://www.sciencedirect.com/science/article/pii/0375960174901686}{Universal
  ratios of critical amplitudes near four dimensions}, Physics Letters A 47~(4)
  (1974) 285--287.
\newblock \href {https://doi.org/https://doi.org/10.1016/0375-9601(74)90168-6}
  {\path{doi:https://doi.org/10.1016/0375-9601(74)90168-6}}.
\newline\urlprefix\url{https://www.sciencedirect.com/science/article/pii/0375960174901686}

\bibitem{1972JETPL..16..178A}
G.~M. {Avdeeva}, A.~A. {Migdal}, {Equation of State in (4 - epsilon) -
  Dimensional Ising Model}, Soviet Journal of Experimental and Theoretical
  Physics Letters 16 (1972) 178.

\bibitem{Onsager:1943jn}
L.~Onsager, {Crystal statistics. 1. A Two-dimensional model with an order
  disorder transition}, Phys. Rev. 65 (1944) 117--149.
\newblock \href {https://doi.org/10.1103/PhysRev.65.117}
  {\path{doi:10.1103/PhysRev.65.117}}.

\bibitem{Wu:1975mw}
T.~T. Wu, B.~M. McCoy, C.~A. Tracy, E.~Barouch, {Spin spin correlation
  functions for the two-dimensional Ising model: Exact theory in the scaling
  region}, Phys. Rev. B 13 (1976) 316--374.
\newblock \href {https://doi.org/10.1103/PhysRevB.13.316}
  {\path{doi:10.1103/PhysRevB.13.316}}.

\bibitem{Zamolodchikov:1987zf}
A.~B. Zamolodchikov, {Integrals of Motion in Scaling Three State Potts Model
  Field Theory}, Int. J. Mod. Phys. A 3 (1988) 743--750.
\newblock \href {https://doi.org/10.1142/S0217751X88000333}
  {\path{doi:10.1142/S0217751X88000333}}.

\bibitem{Zamolodchikov:1989hfa}
A.~B. Zamolodchikov, {Integrable field theory from conformal field theory},
  Adv. Stud. Pure Math. 19 (1989) 641--674.

\bibitem{Yurov:1991my}
V.~P. Yurov, A.~B. Zamolodchikov, {Truncated fermionic space approach to the
  critical 2-D Ising model with magnetic field}, Int. J. Mod. Phys. A 6 (1991)
  4557--4578.
\newblock \href {https://doi.org/10.1142/S0217751X91002161}
  {\path{doi:10.1142/S0217751X91002161}}.

\bibitem{Fateev:1993av}
V.~A. Fateev, {The Exact relations between the coupling constants and the
  masses of particles for the integrable perturbed conformal field theories},
  Phys. Lett. B 324 (1994) 45--51.
\newblock \href {https://doi.org/10.1016/0370-2693(94)00078-6}
  {\path{doi:10.1016/0370-2693(94)00078-6}}.

\bibitem{Xu:2022mmw}
H.-L. Xu, A.~Zamolodchikov, {2D Ising Field Theory in a Magnetic Field: The
  Yang-Lee Singularity} (3 2022).
\newblock \href {http://arxiv.org/abs/2203.11262} {\path{arXiv:2203.11262}}.

\bibitem{Delfino:1997ck}
G.~Delfino, {Universal amplitude ratios in the two-dimensional Ising model},
  Phys. Lett. B 419 (1998) 291--295, [Erratum: Phys.Lett.B 518, 330--330
  (2001)].
\newblock \href {http://arxiv.org/abs/hep-th/9710019}
  {\path{arXiv:hep-th/9710019}}, \href
  {https://doi.org/10.1016/S0370-2693(97)01457-3}
  {\path{doi:10.1016/S0370-2693(97)01457-3}}.

\bibitem{baxter2007exactly}
R.~Baxter, \href{https://books.google.com/books?id=G3owDULfBuEC}{Exactly Solved
  Models in Statistical Mechanics}, Dover books on physics, Dover Publications,
  2007.
\newline\urlprefix\url{https://books.google.com/books?id=G3owDULfBuEC}

\bibitem{Wetterich:1992yh}
C.~Wetterich, {Exact evolution equation for the effective potential}, Phys.
  Lett. B 301 (1993) 90--94.
\newblock \href {http://arxiv.org/abs/1710.05815} {\path{arXiv:1710.05815}},
  \href {https://doi.org/10.1016/0370-2693(93)90726-X}
  {\path{doi:10.1016/0370-2693(93)90726-X}}.

\bibitem{Berges:1995mw}
J.~Berges, N.~Tetradis, C.~Wetterich, {Critical equation of state from the
  average action}, Phys. Rev. Lett. 77 (1996) 873--876.
\newblock \href {http://arxiv.org/abs/hep-th/9507159}
  {\path{arXiv:hep-th/9507159}}, \href
  {https://doi.org/10.1103/PhysRevLett.77.873}
  {\path{doi:10.1103/PhysRevLett.77.873}}.

\bibitem{Delamotte:2007pf}
B.~Delamotte, {An Introduction to the nonperturbative renormalization group},
  Lect. Notes Phys. 852 (2012) 49--132.
\newblock \href {http://arxiv.org/abs/cond-mat/0702365}
  {\path{arXiv:cond-mat/0702365}}, \href
  {https://doi.org/10.1007/978-3-642-27320-9_2}
  {\path{doi:10.1007/978-3-642-27320-9_2}}.

\bibitem{Braun:2011pp}
J.~Braun, {Fermion Interactions and Universal Behavior in Strongly Interacting
  Theories}, J. Phys. G 39 (2012) 033001.
\newblock \href {http://arxiv.org/abs/1108.4449} {\path{arXiv:1108.4449}},
  \href {https://doi.org/10.1088/0954-3899/39/3/033001}
  {\path{doi:10.1088/0954-3899/39/3/033001}}.

\bibitem{Dupuis:2020fhh}
N.~Dupuis, L.~Canet, A.~Eichhorn, W.~Metzner, J.~M. Pawlowski, M.~Tissier,
  N.~Wschebor, {The nonperturbative functional renormalization group and its
  applications}, Phys. Rept. 910 (2021) 1--114.
\newblock \href {http://arxiv.org/abs/2006.04853} {\path{arXiv:2006.04853}},
  \href {https://doi.org/10.1016/j.physrep.2021.01.001}
  {\path{doi:10.1016/j.physrep.2021.01.001}}.

\bibitem{Litim:2001up}
D.~F. Litim, {Optimized renormalization group flows}, Phys. Rev. D 64 (2001)
  105007.
\newblock \href {http://arxiv.org/abs/hep-th/0103195}
  {\path{arXiv:hep-th/0103195}}, \href
  {https://doi.org/10.1103/PhysRevD.64.105007}
  {\path{doi:10.1103/PhysRevD.64.105007}}.

\bibitem{Bohr:2000gp}
O.~Bohr, B.~J. Schaefer, J.~Wambach, {Renormalization group flow equations and
  the phase transition in O(N) models}, Int. J. Mod. Phys. A 16 (2001)
  3823--3852.
\newblock \href {http://arxiv.org/abs/hep-ph/0007098}
  {\path{arXiv:hep-ph/0007098}}, \href
  {https://doi.org/10.1142/S0217751X0100502X}
  {\path{doi:10.1142/S0217751X0100502X}}.

\bibitem{Bervillier:2007rc}
C.~Bervillier, A.~Juttner, D.~F. Litim, {High-accuracy scaling exponents in the
  local potential approximation}, Nucl. Phys. B 783 (2007) 213--226.
\newblock \href {http://arxiv.org/abs/hep-th/0701172}
  {\path{arXiv:hep-th/0701172}}, \href
  {https://doi.org/10.1016/j.nuclphysb.2007.03.036}
  {\path{doi:10.1016/j.nuclphysb.2007.03.036}}.

\bibitem{Braun:2007td}
J.~Braun, B.~Klein, {Scaling functions for the O(4)-model in d=3 dimensions},
  Phys. Rev. D 77 (2008) 096008.
\newblock \href {http://arxiv.org/abs/0712.3574} {\path{arXiv:0712.3574}},
  \href {https://doi.org/10.1103/PhysRevD.77.096008}
  {\path{doi:10.1103/PhysRevD.77.096008}}.

\bibitem{Braun:2009ruy}
J.~Braun, B.~Klein, {Finite-Size Scaling behavior in the O(4)-Model}, Eur.
  Phys. J. C 63 (2009) 443--460.
\newblock \href {http://arxiv.org/abs/0810.0857} {\path{arXiv:0810.0857}},
  \href {https://doi.org/10.1140/epjc/s10052-009-1098-8}
  {\path{doi:10.1140/epjc/s10052-009-1098-8}}.

\bibitem{Benitez:2009xg}
F.~Benitez, J.~P. Blaizot, H.~Chate, B.~Delamotte, R.~Mendez-Galain,
  N.~Wschebor, {Solutions of renormalization group flow equations with full
  momentum dependence}, Phys. Rev. E 80 (2009) 030103.
\newblock \href {http://arxiv.org/abs/0901.0128} {\path{arXiv:0901.0128}},
  \href {https://doi.org/10.1103/PhysRevE.80.030103}
  {\path{doi:10.1103/PhysRevE.80.030103}}.

\bibitem{Stokic:2010piu}
B.~Stokic, B.~Friman, K.~Redlich, {The Functional Renormalization Group and
  O(4) scaling}, Eur. Phys. J. C 67 (2010) 425--438.
\newblock \href {http://arxiv.org/abs/0904.0466} {\path{arXiv:0904.0466}},
  \href {https://doi.org/10.1140/epjc/s10052-010-1310-x}
  {\path{doi:10.1140/epjc/s10052-010-1310-x}}.

\bibitem{Litim:2010tt}
D.~F. Litim, D.~Zappala, {Ising exponents from the functional renormalisation
  group}, Phys. Rev. D 83 (2011) 085009.
\newblock \href {http://arxiv.org/abs/1009.1948} {\path{arXiv:1009.1948}},
  \href {https://doi.org/10.1103/PhysRevD.83.085009}
  {\path{doi:10.1103/PhysRevD.83.085009}}.

\bibitem{Benitez:2011xx}
F.~Benitez, J.~P. Blaizot, H.~Chate, B.~Delamotte, R.~Mendez-Galain,
  N.~Wschebor, {Non-perturbative renormalization group preserving full-momentum
  dependence: implementation and quantitative evaluation}, Phys. Rev. E 85
  (2012) 026707.
\newblock \href {http://arxiv.org/abs/1110.2665} {\path{arXiv:1110.2665}},
  \href {https://doi.org/10.1103/PhysRevE.85.026707}
  {\path{doi:10.1103/PhysRevE.85.026707}}.

\bibitem{Rancon_2013}
A.~Rancon, O.~Kodio, N.~Dupuis, P.~Lecheminant,
  \href{http://dx.doi.org/10.1103/PhysRevE.88.012113}{Thermodynamics in the
  vicinity of a relativistic quantum critical point in2+1dimensions}, Physical
  Review E 88~(1) (Jul 2013).
\newblock \href {https://doi.org/10.1103/physreve.88.012113}
  {\path{doi:10.1103/physreve.88.012113}}.
\newline\urlprefix\url{http://dx.doi.org/10.1103/PhysRevE.88.012113}

\bibitem{Defenu:2014bea}
N.~Defenu, A.~Trombettoni, A.~Codello, {Fixed-point structure and effective
  fractional dimensionality for O$(N)$ models with long-range interactions},
  Phys. Rev. E 92~(5) (2015) 052113.
\newblock \href {http://arxiv.org/abs/1409.8322} {\path{arXiv:1409.8322}},
  \href {https://doi.org/10.1103/PhysRevE.92.052113}
  {\path{doi:10.1103/PhysRevE.92.052113}}.

\bibitem{Codello:2014yfa}
A.~Codello, N.~Defenu, G.~D'Odorico, {Critical exponents of O(N) models in
  fractional dimensions}, Phys. Rev. D 91~(10) (2015) 105003.
\newblock \href {http://arxiv.org/abs/1410.3308} {\path{arXiv:1410.3308}},
  \href {https://doi.org/10.1103/PhysRevD.91.105003}
  {\path{doi:10.1103/PhysRevD.91.105003}}.

\bibitem{Eichhorn:2016hdi}
A.~Eichhorn, L.~Janssen, M.~M. Scherer, {Critical O(N) models above four
  dimensions: Small-N solutions and stability}, Phys. Rev. D 93~(12) (2016)
  125021.
\newblock \href {http://arxiv.org/abs/1604.03561} {\path{arXiv:1604.03561}},
  \href {https://doi.org/10.1103/PhysRevD.93.125021}
  {\path{doi:10.1103/PhysRevD.93.125021}}.

\bibitem{Litim:2016hlb}
D.~F. Litim, E.~Marchais, {Critical $O(N)$ models in the complex field plane},
  Phys. Rev. D 95~(2) (2017) 025026.
\newblock \href {http://arxiv.org/abs/1607.02030} {\path{arXiv:1607.02030}},
  \href {https://doi.org/10.1103/PhysRevD.95.025026}
  {\path{doi:10.1103/PhysRevD.95.025026}}.

\bibitem{Juttner:2017cpr}
A.~J\"uttner, D.~F. Litim, E.~Marchais, {Global Wilson\textendash{}Fisher fixed
  points}, Nucl. Phys. B 921 (2017) 769--795.
\newblock \href {http://arxiv.org/abs/1701.05168} {\path{arXiv:1701.05168}},
  \href {https://doi.org/10.1016/j.nuclphysb.2017.06.010}
  {\path{doi:10.1016/j.nuclphysb.2017.06.010}}.

\bibitem{Roscher:2018ucp}
D.~Roscher, I.~F. Herbut, {Critical $O(2)$ field theory near six dimensions
  beyond one loop}, Phys. Rev. D 97~(11) (2018) 116019.
\newblock \href {http://arxiv.org/abs/1805.01480} {\path{arXiv:1805.01480}},
  \href {https://doi.org/10.1103/PhysRevD.97.116019}
  {\path{doi:10.1103/PhysRevD.97.116019}}.

\bibitem{Yabunaka:2018mju}
S.~Yabunaka, B.~Delamotte, {Why Might the Standard Large $N$ Analysis Fail in
  the O($N$) Model: The Role of Cusps in the Fixed Point Potentials}, Phys.
  Rev. Lett. 121~(23) (2018) 231601.
\newblock \href {http://arxiv.org/abs/1807.04681} {\path{arXiv:1807.04681}},
  \href {https://doi.org/10.1103/PhysRevLett.121.231601}
  {\path{doi:10.1103/PhysRevLett.121.231601}}.

\bibitem{Litim:2000ci}
D.~F. Litim, {Optimization of the exact renormalization group}, Phys. Lett. B
  486 (2000) 92--99.
\newblock \href {http://arxiv.org/abs/hep-th/0005245}
  {\path{arXiv:hep-th/0005245}}, \href
  {https://doi.org/10.1016/S0370-2693(00)00748-6}
  {\path{doi:10.1016/S0370-2693(00)00748-6}}.

\bibitem{DePolsi:2021cmi}
G.~De~Polsi, G.~Hern\'andez-Chifflet, N.~Wschebor, {Precision calculation of
  universal amplitude ratios in O(N) universality classes: Derivative expansion
  results at order O(\ensuremath{\partial}4)}, Phys. Rev. E 104~(6) (2021)
  064101.
\newblock \href {http://arxiv.org/abs/2109.14731} {\path{arXiv:2109.14731}},
  \href {https://doi.org/10.1103/PhysRevE.104.064101}
  {\path{doi:10.1103/PhysRevE.104.064101}}.

\bibitem{Macfarlane:1974vp}
A.~J. Macfarlane, G.~Woo, {$\phi^3$ Theory in Six Dimensions and the
  Renormalization Group}, Nucl. Phys. B 77 (1974) 91--108, [Erratum:
  Nucl.Phys.B 86, 548--548 (1975)].
\newblock \href {https://doi.org/10.1016/0550-3213(74)90306-X}
  {\path{doi:10.1016/0550-3213(74)90306-X}}.

\bibitem{Kurtze:1979zz}
D.~A. Kurtze, M.~E. Fisher, {Yang-Lee edge singularities at high temperatures},
  Phys. Rev. B 20 (1979) 2785--2796.
\newblock \href {https://doi.org/10.1103/PhysRevB.20.2785}
  {\path{doi:10.1103/PhysRevB.20.2785}}.

\bibitem{Tetradis:1993ts}
N.~Tetradis, C.~Wetterich, {Critical exponents from effective average action},
  Nucl. Phys. B 422 (1994) 541--592.
\newblock \href {http://arxiv.org/abs/hep-ph/9308214}
  {\path{arXiv:hep-ph/9308214}}, \href
  {https://doi.org/10.1016/0550-3213(94)90446-4}
  {\path{doi:10.1016/0550-3213(94)90446-4}}.

\bibitem{Kos:2016ysd}
F.~Kos, D.~Poland, D.~Simmons-Duffin, A.~Vichi, {Precision Islands in the Ising
  and $O(N)$ Models}, JHEP 08 (2016) 036.
\newblock \href {http://arxiv.org/abs/1603.04436} {\path{arXiv:1603.04436}},
  \href {https://doi.org/10.1007/JHEP08(2016)036}
  {\path{doi:10.1007/JHEP08(2016)036}}.

\bibitem{Roberge:1986mm}
A.~Roberge, N.~Weiss, {Gauge Theories With Imaginary Chemical Potential and the
  Phases of {QCD}}, Nucl. Phys. B 275 (1986) 734--745.
\newblock \href {https://doi.org/10.1016/0550-3213(86)90582-1}
  {\path{doi:10.1016/0550-3213(86)90582-1}}.

\bibitem{Singh:2021pog}
S.~Singh, P.~Dimopoulos, L.~Dini, F.~Di~Renzo, J.~Goswami, G.~Nicotra,
  C.~Schmidt, K.~Zambello, F.~Ziesche, {Lee-Yang edge singularities in lattice
  QCD : A systematic study of singularities in the complex $\mu_B$ plane using
  rational approximations}, in: {38th International Symposium on Lattice Field
  Theory}, 2021.
\newblock \href {http://arxiv.org/abs/2111.06241} {\path{arXiv:2111.06241}}.

\end{thebibliography}

\end{document}